\documentclass[a4paper,11pt]{article}
\usepackage{jcappub} 
\usepackage{lineno}
\usepackage{mathtools}
\usepackage[normalem]{ulem}
\usepackage{macros}
\graphicspath{{figure/}}

\usepackage[utf8]{inputenc}
\usepackage[T1]{fontenc}

\arxivnumber{2311.07557}
\title{\boldmath Backreaction of axion-SU(2) dynamics during inflation}

 \author[a,b]{Oksana~Iarygina,}
 \author[c,d]{Evangelos~I.~Sfakianakis,}
 \author[a,b]{Ramkishor~Sharma,}
 \author[a,e,f,g]{and~Axel~Brandenburg}
 \affiliation[a]{Nordita, KTH Royal Institute of Technology and Stockholm University, Hannes Alfv\'ens v\"ag 12, 10691 Stockholm, Sweden}
\affiliation[b]{The Oskar Klein Centre, Department of Physics, Stockholm University, AlbaNova, 10691 Stockholm, Sweden}
\affiliation[c]{Department of Physics, Case Western Reserve University,
10900 Euclid Avenue, Cleveland, OH 44106, USA}
\affiliation[d]{Institut de Fisica d'Altes Energies (IFAE), The Barcelona Institute of
Science and Technology (BIST), Campus UAB, 08193 Bellaterra, Barcelona, Spain}
\affiliation[e]{The Oskar Klein Centre, Department of Astronomy, Stockholm University, 10691 Stockholm, Sweden}
\affiliation[f]{McWilliams Center for Cosmology \& Department of Physics, Carnegie Mellon University, Pittsburgh, PA 15213, USA}
\affiliation[g]{School of Natural Sciences and Medicine, Ilia State University, 3-5 Cholokashvili Avenue, 0194 Tbilisi, Georgia}

\emailAdd{oksana.iarygina@su.se}
\emailAdd{esfakianakis@ifae.es}
\emailAdd{ramkishor.sharma@su.se}
\emailAdd{brandenb@nordita.org}

\abstract{

We consider the effects of backreaction on axion-SU(2)
dynamics during inflation.
We use the linear evolution equations for the gauge field modes and compute their backreaction on the background quantities numerically using the Hartree approximation.
We show that the spectator chromo-natural inflation attractor is unstable when back-reaction becomes important.
Working within the constraints of the linear mode equations,
we find a new dynamical attractor solution for the axion field and the vacuum expectation
value of the gauge field, where the latter has an opposite sign
with respect to the chromo-natural inflation solution.
Our findings are of particular interest to the phenomenology of
axion-SU(2) inflation,
as they demonstrate the instability of the usual trajectory due to large backreaction effects.
The viable parameter space of the model becomes significantly altered, provided future non-Abelian lattice simulations confirm the existence of the new dynamical attractor.
In addition, the backreaction effects lead to characteristic oscillatory
features in the primordial gravitational wave background that are potentially
detectable with upcoming gravitational wave detectors.
}

\makeatletter
\gdef\@fpheader{}
\makeatother

\begin{document}
\maketitle
\flushbottom

\section{Introduction}

Natural inflation was originally proposed in order to address the UV
sensitivity problem of inflation \cite{Freese:1990rb, Adams:1992bn}.
By identifying the inflaton with an axion, the pseudo-scalar field
possesses a shift symmetry that protects the inflationary potential against radiative corrections.
The inflationary potential is generated through instanton effects and
in the early versions of natural inflation the potential is a sinusoidal function of the field.

The shift symmetry also dictates the possible couplings of the inflaton to other fields,
since an axion can only couple derivatively to other fields.
In particular, the lowest order coupling to fermions $\psi$ and gauge bosons $A_\mu$ are
${ f^{-1}} (\partial_\mu\phi)\bar\psi \gamma_\mu \gamma^5 \psi$ and ${ f^{-1}}\phi F_{\mu\nu}\tilde F^{\mu\nu}$,
respectively, where $F_{\mu\nu}$ is the  field strength tensor and 
$\tilde F_{\mu\nu} = \epsilon^{\mu\nu\alpha\beta}F_{\alpha\beta}$ is the dual field strength tensor
(see e.g.\ \cite{ Adshead:2015kza, Adshead:2021ezw, Ferreira:2014zia, Adshead:2015pva}).

The coupling of an axion inflaton to gauge fields through a Chern-Simons
term has attracted significant attention in the literature.
In the case of an Abelian field, the parity-violating nature of the
coupling leads to the two helicities developing a different effective frequency.
One of them can even become tachyonic, when the velocity of the inflaton is high enough.
After the end of inflation, tachyonic production of gauge fields can lead to instantaneous preheating \cite{Adshead:2015pva}.
Identifying the gauge field with the hypercharge sector of the Standard Model can lead to the generation of observationally relevant cosmological magnetic fields \cite{Fujita:2015iga, Adshead:2016iae}.
During inflation, the production of gauge fields can lead to observable non-Gaussianity.

Depending on the axion-gauge coupling strength, the tachyonic amplification of the gauge fields can arise during inflation.
In this case, the generation of gauge fields leads to a significant backreaction on the inflaton, leading to a sudden drop of its velocity.
Once the gauge fields are diluted by the expansion of space-time, the backreaction term subsides and the inflaton starts rolling again.
This can lead to periodic bursts of gauge field production during
inflation \cite{Cheng:2015oqa, Notari:2016npn, Sobol:2019xls, DallAgata:2019yrr, Domcke:2020zez, Peloso:2022ovc, Caravano:2022epk, Garcia-Bellido:2023ser, vonEckardstein:2023gwk}, as has been shown analytically and numerically.
However, recent lattice simulations showed that the inclusion of inhomogeneous
backreaction and a larger dynamical range can significantly change the
resulting dynamics \cite{Figueroa:2023oxc}.

Despite the interesting backreaction dynamics that occurs for large
axion-gauge coupling, the rolling of the axion in the linear regime is
determined by the potential and Hubble friction.
By replacing the Abelian field with a non-Abelian one, this ceases to be true.
The fact that SU(2) fields (see ref.~\cite{Fujita:2021eue} for the
generalization to SU(N) fields) possess a non-trivial vacuum structure
leads to a new inflationary attractor, in which the dominant source of
friction for the rolling axion is not the Hubble term, but the non-Abelian
field VEV \cite{Maleknejad+11a,Maleknejad+11b,Adshead:2012kp,  Adshead:2013qp, Adshead:2013nka}.

This family of models, collectively named chromo-natural inflation,
allows for slow-roll inflation even in steep potentials.
Due to the parity-violating Chern-Simons coupling, one of the tensor modes
 of the SU(2) sector experiences a similar instability to one of the helicities in the Abelian case. 
 However, the fact that the SU(2) tensor mode is linearly coupled to the gravitational sector leads to a similar enhancement of chiral gravitational waves (GWs).
 
While chromo-natural inflation with a cosine potential has been shown to be 
incompatible with CMB observations \cite{Adshead:2013nka}, spontaneous breaking of the 
 SU(2) symmetry or going beyond the cosine potential can
 bring the model in agreement with current data \cite{Adshead:2016omu, Maleknejad:2016qjz}. A further generalization of CNI
 was proposed in ref.~\cite{Dimastrogiovanni:2016fuu}, where the axion-SU(2) action was treated as a spectator sector. 
This separates the inflationary sector, which is responsible for scalar
fluctuations, from the chromo-natural sector, which can produce detectable
B-modes, while remaining subdominant in both scalar fluctuations and
energy density during inflation.
This family of models can be described as ``spectator chromo-natural inflation'' (SCNI) and their GW spectra are directly related to the shape of the axion potential \cite{Fujita:2018ndp}. Currently, the dynamics of spectator non-Abelian gauge fields during inflation is an active research direction 
\cite{Sheikh-Jabbari:2012tom, Maleknejad:2012wqk, Fujita:2017jwq, Caldwell:2017chz, Agrawal:2017awz, Lozanov:2018kpk, Dimastrogiovanni:2018xnn, Domcke:2018rvv, Fujita:2018vmv, Mirzagholi:2019jeb, Maleknejad:2019hdr, Wolfson:2020fqz, Mirzagholi:2020irt, Iarygina:2021bxq, Fujita:2021flu, Fujita:2022fff, Adshead:2022ecl, Bagherian:2022mau, Dimastrogiovanni:2023oid}.

 Previously, the effects of backreaction in spectator axion-SU(2) inflation were 
estimated by comparing the backreaction contributions to other terms in equations of motion \cite{Dimastrogiovanni:2016fuu, Fujita:2017jwq, Maleknejad:2018nxz, Papageorgiou:2019ecb},
or more generally in ref.~\cite{Ishiwata:2021yne} for the case when the backreaction integral is regularized\footnote{In the slow-roll approximation, the backreaction integrals of gauge field tensor perturbations onto background quantities 
can be
expressed via linear combinations of the Whittaker functions that are divergent and require regularization \cite{Maleknejad:2018nxz}.}. The authors of the ref.~\cite{Ishiwata:2021yne} studied the stability of slow-roll dynamics during axion-SU(2) inflation and 
found that the slow-roll solutions become unstable when the backreaction terms dominate in the equations of motion.
In this work, we go beyond previous studies of axion-SU(2) dynamics during inflation by considering the effects of backreaction  without the regularization of backreaction integrals. 
We use the linear equations of motion for the tensor SU(2) fluctuations and self-consistently solve the background equations for the axion and SU(2) VEV, including the homogeneous (averaged) backreaction from tensor fluctuations. This is in spirit similar to the analysis performed in ref.~\cite{Garcia-Bellido:2023ser} for the Abelian case. We must note that ref.~\cite{Caravano:2022epk} largely validated these calculations, while more recent simulations~\cite{Figueroa:2023oxc} point out the importance of inhomogeneous effects during the strong backreaction regime. 
Our analysis can therefore be considered an important and necessary  next step into the uncharted backreaction regime of axion-SU(2) dynamics during inflation.
Our use of the linearized equations of motion for the gauge fields does not allow us to capture possible non-Abelian effects.
While we can safely conclude that backreaction leads to the destabilization of the ``standard'' chromo-natural attractor, full lattice simulations are required to validate the existence of the
new dynamical late-time attractor described below.
This is left for future work.

Our manuscript is organized as follows.
In section~\ref{sec:review} we review spectator chromo-natural inflation
and provide the necessary equations and analytical solutions.
The numerical procedure is described in section~\ref{sec:numerics},
followed by the results and semi-analytical analysis of the solution.
We conclude in section~\ref{sec:summary}.
In \App{SetOfParameters} we present the set of parameters used for the numerical computations.
In \App{Artifacts} we discuss possible numerical artifacts.
In \App{appendix_Stages} we give details of the backreaction dynamics and
in \App{ASsolution} discuss the behavior of tensor perturbations on the dynamical attractor solution found in this work
and provide a comparison with axion-U(1) inflation.

\section{Review of spectator axion-SU(2) inflation} 
\label{sec:review}

In this section, we review the spectator axion-SU(2) inflation or the spectator chromo-natural inflation model,
outlining the background and perturbation analysis based on previous works
\cite{Adshead:2012kp, Adshead:2013nka, Dimastrogiovanni:2016fuu}.

\subsection{Model and background evolution}
The action for spectator axion-SU(2) inflation is given by \cite{Dimastrogiovanni:2016fuu}
\begin{equation}\label{spectatorAction}
S=\int d^4 x \sqrt{-\text{det} \, g_{\mu\nu}}\left[ \frac{M_{\text{pl}}^2}{2}R -\frac{1}{2}(\partial \phi)^2-V(\phi)-\frac{1}{2}(\partial \chi)^2-U(\chi)
-\frac{1}{4}F^a_{\mu\nu}F^{a\, \mu\nu}+\frac{\lambda \chi}{4f}F^a_{\mu\nu}\tilde{F}^{a\, \mu\nu} \right],
\end{equation}
where $R$ is the space-time Ricci scalar, $\phi(t)$ and $V(\phi)$ are the inflaton field and its potential, respectively,
$\chi(t)$ and $U(\chi)$ are the axion field and its potential,
$F^a_{\mu\nu}=\partial_{\mu} A^a_{\nu}-\partial_{\nu} A^a_{\mu}-g \epsilon^{abc}A^b_{\mu}A^c_{\nu}$
is the field strength of the SU(2) gauge field
$A^a_{\mu}$, $\tilde{F}^{a\, \mu\nu}=\epsilon^{\mu\nu\rho\sigma}F^a_{\rho\sigma}/\left(2\sqrt{-\text{det} \, g_{\mu\nu}}\, \right)$
is its dual ($\epsilon^{\mu \nu\alpha\beta}$ is the antisymmetric tensor and $\epsilon^{0123}=1$), $g$ is the gauge
field coupling, $\lambda$ is the coupling constant between the gauge
and axion sectors, $f$ is the axion decay constant, and $M_\text{pl}$ is the reduced Planck mass.

In this work we use the axion potential of the form
\begin{equation}
    U(\chi)=\mu^4\left(1+\cos \frac{\chi}{f} \right),
\end{equation}
where $\mu$ is a constant that sets the energy scale of the axion field.
In this convention, the axion field takes values $\chi \in \left[0, \pi f \right]$.
The potential for the inflation field, $V(\phi)$, is left unspecified. 

We work with the FLRW metric
\begin{equation}
ds^2=-dt^2+a(t)^2 \delta_{ij}dx^i dx^j,
\end{equation}
where $i,j$ indicate the spatial directions.
An isotropic solution for the background is given by the gauge field configuration \cite{Maleknejad+11a,Maleknejad+11b} 
\begin{gather}
A^a_0=0,\qquad
A^a_i=\delta ^a_i a(t)Q(t), \label{ansatzA}
\end{gather}
which is also an attractor \cite{Maleknejad:2013npa}.
For this ansatz, the closed system of equations for the vacuum expectation value (VEV) of the gauge field $Q(t)$ and the Hubble parameter $H(t)$ is given by
\begin{gather}
M_{\text{pl}}^2\dot{H}=-\frac{1}{2}\dot{\phi}^2-\frac{1}{2}\dot{\chi}^2-\left[ ( \dot{Q}+H Q)^2+ g^2 Q^4\right], \label{eom1} \\
3 M_{\text{pl}}^2 H^2=\frac{\dot{\phi}^2}{2}+V(\phi)+\frac{\dot{\chi}^2}{2}+U(\chi)+\frac{3}{2}\left(\dot{Q}+H Q\right)^2+\frac{3}{2}g^2 Q^4, \\
 \ddot{Q}+3H\dot{Q}+\left(\dot{H}+2H^2\right)Q+2g^2Q^3=\frac{g\lambda }{f}\dot{\chi}Q^2,\\
 \ddot{\chi}+3H\dot{\chi}+U_{\chi}(\chi)=-\frac{3g\lambda}{f}Q^2\left( \dot{Q}+HQ\right),\\
 \ddot{\phi}+3H\dot{\phi}+V_{\phi}(\phi)=0, \label{eom5}
\end{gather}
where $V_{\phi}(\phi)=\partial V(\phi)/\partial \phi$, $V_{\chi}(\chi)=\partial U(\chi)/\partial \chi$, and an overdot denotes a derivative with respect to cosmic time $t$.
The Hubble slow-roll parameters are defined as
\begin{equation}
    \epsilon_H=-\frac{\dot{H}}{H^2}, \quad \eta_H=-\frac{\ddot{H}}{2H\dot{H}},
\end{equation}
which are much smaller than unity during inflation. The first slow-roll parameter $\epsilon_H$ contains contributions from the inflaton field $\phi$ and the spectator sector that consists of the axion and gauge fields 
\begin{equation}
    \epsilon_H=\epsilon_{\phi}+\epsilon_{Q_E}+\epsilon_{Q_B}+\epsilon_{\chi}.
\end{equation}
The various contributions are defined as
\begin{gather}
\epsilon_{\phi}=\frac{\dot{\phi}^2}{2M_{\text{pl}}^2 H^2},\quad \epsilon_{Q_E}=\frac{( \dot{Q}+H Q)^2}{M_{\text{pl}}^2 H^2},\quad \epsilon_{Q_B}=\frac{g^2Q^4}{M_{\text{pl}}^2 H^2},\quad
\epsilon_{\chi}=\frac{\dot{\chi}^2}{2M_{\text{pl}}^2 H^2}.
\end{gather}
For the axion-gauge sector to remain a spectator, their energy densities must be subdominant to that of the inflaton, i.e.,
\begin{equation}
    \rho_{\phi}\gg \rho_{\chi},\,\rho_{Q_{E}},\,\rho_{Q_B},
\end{equation}
where the energy densities are given by
\begin{eqnarray}
\rho_{\phi}=\frac{1}{2}\dot{\phi}^2+V(\phi), \quad
\rho_{\chi}=\frac{1}{2}\dot{\chi}^2+U(\chi), \quad
\rho_{Q_E}=\frac{3}{2}( \dot{Q}+H Q)^2,\quad
\rho_{Q_B}=\frac{3}{2}g^2 Q^4.
\end{eqnarray}

The original chromo-natural inflation model in the slow-roll approximation has an attractor solution \cite{Adshead:2012kp, Adshead:2013nka}
\begin{equation}
    \begin{split}
 \frac{\lambda}{f}\dot{\chi}=2gQ+\frac{2H^2}{gQ},\qquad
 \dot{Q}=-HQ+\frac{f}{3g\lambda}\frac{U_{\chi}}{Q^2}. \label{attractorCNI}
\end{split}
\end{equation}
The VEV of the gauge field that minimizes
the axion effective potential is 
\begin{equation}
     Q\simeq \left(\frac{-f U_{\chi}(\chi)}{3 g \lambda H} \right)^{1/3},
\end{equation}
which is a solution of \Eq{attractorCNI} when $Q$ is constant.
It is convenient to introduce the parameters
\begin{equation}\label{mQxidef}
    m_Q=\frac{g Q}{H}, \qquad \xi=\frac{\lambda}{2fH}\dot{\chi},
\end{equation}
where the dimensionless mass parameter $m_Q$ characterizes the mass of the gauge field 
fluctuations and controls their amplification. On the chromo-natural inflation attractor, $m_Q$ and $\xi$ are related via $\xi \simeq m_Q+1/m_Q$.

\subsection{Perturbations}
Let us now review the perturbations in the spectator axion-SU(2) model.
We adopt the gauge choice and decomposition for field fluctuations following ref.~\cite{Dimastrogiovanni:2012ew} of the form
\begin{equation}
    \begin{split}
     \phi &= \phi+\delta\phi,\\
        \chi &= \chi+\delta\chi,\\
      A^1_{\mu}  &= a\,(Y_1,\,Q+\delta Q+t_+,\,t_{\times},\, \partial_z M_1),\\
      A^2_{\mu}  &= a\,(Y_2,\,t_{\times},\,Q+\delta Q-t_+,\, \partial_z M_2),\\
       A^3_{\mu}  &= a\,(Y_z,\,0,\,0,\,Q+\delta Q+\partial_z\partial_z M),
    \end{split}
\end{equation}
together with
\begin{equation}
    g_{\mu\nu}=a^2
\begin{pmatrix}
  -1+2\varphi  & B_1 & B_2 & \partial_z B\\ 
  {}           & 1+h_{+} & h_{\times} & 0\\
  {}           & {}      &  1-h_{+}   & 0\\
  {}           &{}       &{}          &1 
\end{pmatrix}
.
\end{equation}
The perturbations consist of seven scalar modes $(\delta \phi,\, \delta \chi,\, Y,\, \delta Q,\, M,\, \varphi,\, B )$, six vector modes $(Y_{1,2},\, M_{1,2},\, B_{1,2})$ and four tensor modes $(t_+,\, t_{\times},\, h_+,\, h_{\times})$. At the linear level, all perturbations are decoupled from each other. Vector perturbations decay on superhorizon scales and at the linear level metric fluctuations can be neglected \cite{Dimastrogiovanni:2012ew}.

The scalar perturbations have been studied in detail at the linear \cite{Dimastrogiovanni:2016fuu} and nonlinear levels \cite{Papageorgiou:2018rfx, Papageorgiou:2019ecb}.
It was shown that for $m_Q< \sqrt{2}$, the scalar perturbations are tachyonically unstable \cite{Dimastrogiovanni:2012ew}.
The combination from linear and nonlinear analyses leads to the constraint
\cite{Papageorgiou:2019ecb}
\begin{equation}
    \sqrt{2}<m_Q\leq \left( \frac{g^2}{32 \pi^2 \, P_{\zeta,\rm{CMB}}}\right)^{1/4} \simeq 35 \sqrt{g}, \label{constraintScalar}
\end{equation}
on the parameter $m_Q$, where $P_{\zeta,\rm{CMB}}=2.1 \cdot 10^{-9}$.

Let us now turn to the discussion of tensor perturbations.
It is convenient to express the plus and cross polarizations of tensor perturbations via the left-handed and right-handed polarizations as
\begin{gather}\label{LRmodes}
h_{+}=\frac{h_L+h_R}{\sqrt{2}}, \quad h_{\times}=\frac{h_L-h_R}{i\sqrt{2}},\quad t_{+}=\frac{t_L+t_R}{\sqrt{2}},\quad t_{\times}=\frac{t_L-t_R}{i\sqrt{2}}.
\end{gather}
We canonically normalize them by introducing
\begin{equation}
h_{L,R}=\frac{\sqrt{2}}{M_p a}\psi_{L,R},\qquad t_{L,R}=\frac{1}{\sqrt{2}a}T_{L,R}.
\end{equation}
We will work with conformal time defined as
\begin{equation}
    \eta=\int_0^t \frac{dt}{a(t)},
\end{equation}
which, with the near de Sitter expansion, leads to
\begin{equation}
    a=-\frac{1}{H\eta}
\end{equation}
for the scale factor.
In the following, derivatives with respect to $\eta$ are denoted by primes.
In conformal time and to leading order in slow-roll,
the equations of motion for the tensor perturbations are 
\begin{align}
\psi_{R,L}''+\left(k^2-\frac{2}{\eta^2}\right)\psi_{R,L}&=\frac{2 \sqrt{\epsilon_{Q_E}}}{\eta} T_{R,L}'+\frac{2 \sqrt{\epsilon_{Q_B}}}{\eta^2}(m_Q\pm k\eta)T_{R,L}, \label{psieqn}\\
T_{R,L}''+\left\{k^2+\frac{2}{\eta^2}\big[m_Q \xi\pm k \eta(m_Q+\xi)\big]\right\}T_{R,L}&=-\frac{2 \sqrt{\epsilon_{Q_E}}}{\eta} \psi_{R,L}'\nonumber\\
        &+\frac{2 }{\eta^2}\left[\sqrt{\epsilon_{Q_B}}(m_Q\pm k\eta)+\sqrt{\epsilon_{Q_E}}\right]\psi_{R,L}.\label{treqn}
\end{align}
Here $k$ is the wave number.
The spectator axion-SU(2) model is known to have a transient growth of one of the polarizations of the gauge field tensor modes that leads
to the production of chiral GWs.
The produced GW background is enhanced with respect to the
standard single-field slow-roll models of inflation with predictions
potentially observable by near-future B-mode experiments.

The total GW power spectrum is defined as
\begin{equation}
    \sum_{i,j}\langle h_{ij}(\kk)h_{ij}(\kk')\rangle=(2\pi)^3\delta^3(\kk+\kk')\,{\cal P}_h^{\rm tot}(k),
\end{equation}
where ${\cal P}_h^{\rm tot}(k)$ can be expressed in terms of left and right polarization modes as
\begin{equation}\label{pheqn}
{\cal P}_h^{\rm tot}(k)=2{\cal P}_h^{L}(k)+2{\cal P}_h^{R}(k),
\end{equation}
where ${\cal P}_h^{(s)}$ is
the late-time sourced GW power spectrum, defined as
\begin{equation}
{\cal P}_h^{(s)}(k)=\frac{H^2}{\pi^2 M_\text{pl}^2}\Bigl|\sqrt{2k}\,\left(-\frac{k}{a H}\right) \lim_{\eta\rightarrow 0} \psi_R^{(s)}(k,\eta)\Bigr|^2.
\end{equation}
It is convenient to introduce a parameter that characterizes the enhancement of the GW background with respect to the vacuum prediction,
\begin{equation}
    {\cal R}_{\rm GW}=\frac{{\cal P}_h^{(s)}}{{\cal P}_h^{(v)}},
\end{equation}
where the vacuum prediction for the power spectrum is given by
\begin{equation}
    {\cal P}_h^{(v)}(k)=\frac{2 H^2}{\pi^2 M_\text{pl}^2}.
\end{equation}
In order to have sizable GW production, the parameter range of the model has to be chosen such that ${\cal R}_{\rm GW}\gtrsim 1$. This requirement leads to the constraint \cite{Papageorgiou:2019ecb}
\begin{equation}
    g \lesssim 1.8 \cdot 10^{-5} \, m_Q^2 \, e^{1.8 m_Q}. \label{constraintRGW}
\end{equation}

The growth of tensor modes results in the backreaction \cite{Dimastrogiovanni:2016fuu, Fujita:2017jwq,  Maleknejad:2018nxz,  Papageorgiou:2019ecb, Ishiwata:2021yne} on the background equations of motion  \eqref{eom1}--\eqref{eom5}.
Taking into account the contribution from backreaction, the background equations of motion in conformal time take the form 
\begin{gather}
Q''+2 \mathcal{H}Q'+\left(\mathcal{H}'+\mathcal{H}^2\right)Q+2g^2 a^2 Q^3-\frac{g\lambda }{f}a\chi'Q^2+a^2 {\cal T}^Q_{\rm BR}=0,\label{qeqn}\\
\chi''+2\mathcal{H}\chi'+a^2U_{\chi}(\chi)+\frac{3g\lambda}{f}aQ^2\left(Q'+\mathcal{H}Q\right)+a^2 {\cal T}^{\chi}_{\rm BR}=0,\label{chieqn}
\end{gather}
with $\mathcal{H}=a'/a$.
The backreaction terms are evaluated in the Hartree approximation, leading to the integrals over mode functions
\begin{gather}
 {\cal T}^Q_{\rm BR}=\frac{g}{3a^2}\int \frac{d^3 k}{(2\pi)^3}\left( \xi H -\frac{k}{a}\right)|T_R|^2,\label{qbackint}\\
 {\cal T}^{\chi}_{\rm BR}=-\frac{\lambda}{2 a^4 f}\frac{d}{d\eta}\int  \frac{d^3 k}{(2\pi)^3}\left(a\, m_Q H-k\right)|T_R|^2.\label{chibackint}
\end{gather}
It is worth noting that in this work we consider \textit{homogeneous} backreaction, where the spatial gradients of inflation and axion fields are neglected, keeping these fields homogeneous during inflation.

\begin{figure}
\centering
 \includegraphics[width=0.55\textwidth]{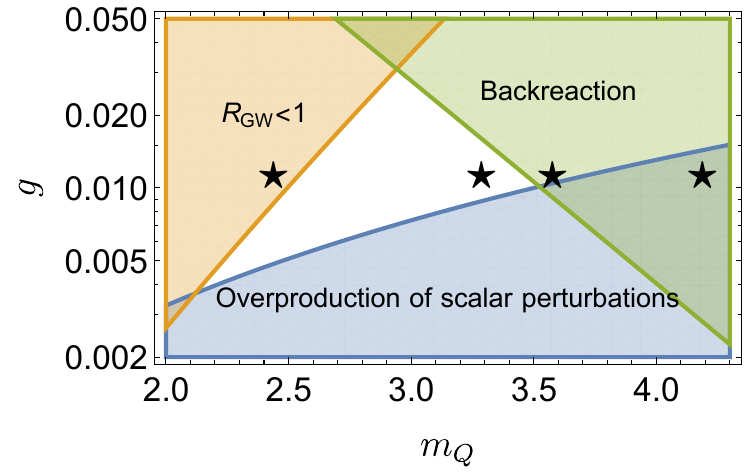}
\caption{Constraints on the spectator axion-SU(2) model similar as in ref.~\cite{Papageorgiou:2019ecb} with indication of parameters used in the current work.
The stars correspond to the parameters (from left to right) $m_Q=2.44$, $3.29$, $3.58$, $4.19$ with $g=0.011$
(runs $\mu1$, $\mu3$, $\mu4$, and $\mu5$ from table~\ref{Tsummary}).
}\label{Fig:RegionPlot}
\end{figure}

In the small-backreaction regime with an approximately constant $m_Q$ parameter, the spectator axion-SU(2) model can be solved analytically \cite{Adshead:2013qp}. The regime of small backreaction is achieved with the constraint \cite{Papageorgiou:2019ecb}
\begin{equation}
    g \ll \left( \frac{24\pi^2}{2.3\cdot e^{3.9 m_Q}}\frac{1}{1+m_Q^{-2}}\right)^{1/2}. \label{constraintBR}
\end{equation}
In figure~\ref{Fig:RegionPlot}, we plot the three constraints \eqref{constraintScalar}, \eqref{constraintRGW}, and \eqref{constraintBR}\footnote{The boundaries of different regions correspond to equality signs in the corresponding equations.} on the parameter range of the theory,
indicating the fiducial parameters used in the current work.

\section{Numerical treatment of the backreaction}
\label{sec:numerics}

To study the backreaction in axion-SU(2) inflation, we now
solve the perturbation equations along with those for the background numerically.
We begin by describing the numerical method and then discuss the results.

\subsection{Numerical implementation}

We solve \Eqs{treqn}{psieqn} for both the left- and right-handed components of $T_{R,L}$ and $\psi_{R,L}$.
For each perturbation variable, we solve the equations for the real and imaginary parts and
represent them on a $k$ mesh to compute the integrals in \eqs{qbackint}{chibackint}.
For most of our studies, we use the logarithmic wave number along with conformal time
as the independent coordinates.
In that case, we use $n_k$ points in $\ln k$ that are separated
by uniform intervals in $\ln k$ in the range
\begin{equation}
n_{\min}\leq\ln(k/a_0 H) \leq n_{\max}.
\end{equation}

To solve the background \eqs{qeqn}{chieqn}, we compute the integrals \eqs{qbackint}{chibackint} up to second-order accuracy.
We advance the solution in conformal time using a third-order time-stepping scheme.
The initial conformal time is $\eta_i$ and the final one is $\eta_f$.
In practice, we choose $\eta_i=-\mathcal{H}_i^{-1}$ with
$\mathcal{H}_i=a_i H$ and $a_i=1$ along with $\eta_f=-10^{-5}$,
which corresponds to a total duration of $N\approx25$ $e$-folds.
The length of the conformal time step is then usually chosen to be $\Delta\eta=10^{-6}/\mathcal{H}$.

It is convenient to use the compute and data management infrastructure
provided by the {\sc Pencil Code} \cite{JOSS}, which allows for efficient
parallelization using the Message Passing Interface library.
In some exploratory cases, we also solved the equations on a mesh
where $n_{\min}$ and $n_{\max}$
grow in time such that the main contributions
to the integral are captured during the entire evolution.

We initialize the perturbation variables with the Bunch-Davies initial condition. Specifically, we set the initial conditions for the real and imaginary parts of the perturbation variables as follows:
\begin{align}
    T_{R,L}=\frac{e^{-ik\eta}}{\sqrt{2k}}, \quad 
    T_{R,L}'=-ik\, T_{R,L}.
\end{align}
The same initial conditions are used for $\psi_{R,L}$. To discard the contributions from quantum vacuum fluctuations of $T_{R,L}$ in the calculation of the integrals in \eqs{qbackint}{chibackint}, we use the criterion
that for wave numbers where $|T_{R,L}|^2<1/2k$, the value of $|T_{R,L}|^2$ is replaced by zero.

\subsection{New late-time attractor solution}
\label{sec:results}

\begin{figure*}
\begin{center}
\includegraphics[scale=0.6]{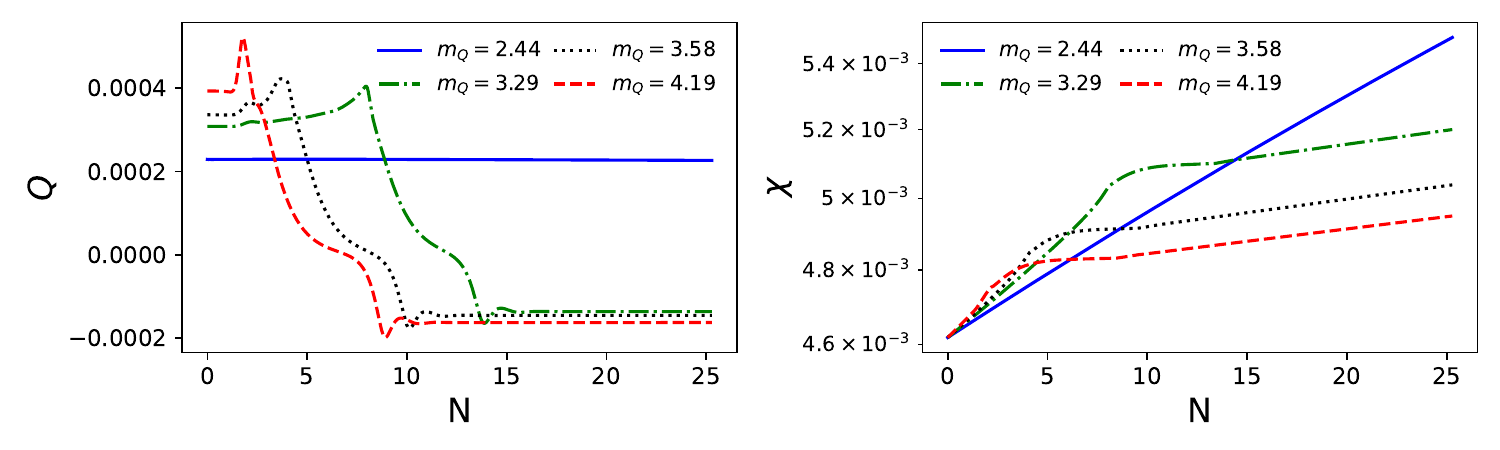}
\end{center}\caption[]{Evolution of $Q$ and $\chi$ with $N$ is shown for different initial parameters.
The blue solid, green dot-dashed, black dotted, and  red dashed
curves correspond to runs $\mu$1,  $\mu$3,  $\mu$4,
and $\mu$5 respectively from table~\ref{Tsummary} with $g=0.011$ and
$m_Q=2.44,\, 3.29,\,3.58,\, 4.19$ respectively.
For the blue curve, the backreaction is negligible therefore the value of $Q$ remains constant.
However, for a larger value of $m_Q$ (green, black and red curves), the backreaction effects are important.
Using equation~\eqref{eqstage30} from section~\ref{sec:semianalytical}, the velocity of $\chi$ is $d\chi/dN = -2Hf/(g\lambda Q)$,
which can be evaluated using the data of table~\ref{Tsummary}, leading to $d\chi/dN = -8.2 \times 10^{-6}$, $-7.7 \times 10^{-6}$, and $-6.9 \times 10^{-6}$
for the green dot-dashed, black dotted and red dashed curves respectively.
These estimates agree with the $\chi$ evolution shown in this figure.
}\label{Q_and_chi_plot}\end{figure*}

We have performed a range of simulations with different values of
$\mu$, $g$, and $\lambda$; see \App{SetOfParameters} for a summary.
In \Fig{Q_and_chi_plot}, we show the time evolution of $Q$ and $\chi$ for runs~$\mu$1, $\mu$3, $\mu$4, and $\mu$5,
which corresponds to $m_{Q}=2.44$, $3.29$, $3.58$, and $4.19$, respectively. When $m_Q$ is large enough, the backreaction of the perturbations becomes important and the system
undergoes a transition to the new dynamical attractor with negative
values of $Q$ and a reduced velocity for $\chi$ after about $N=2$--$10$ $e$-folds.

\begin{figure*}\begin{center}
\includegraphics[width=.95\textwidth]{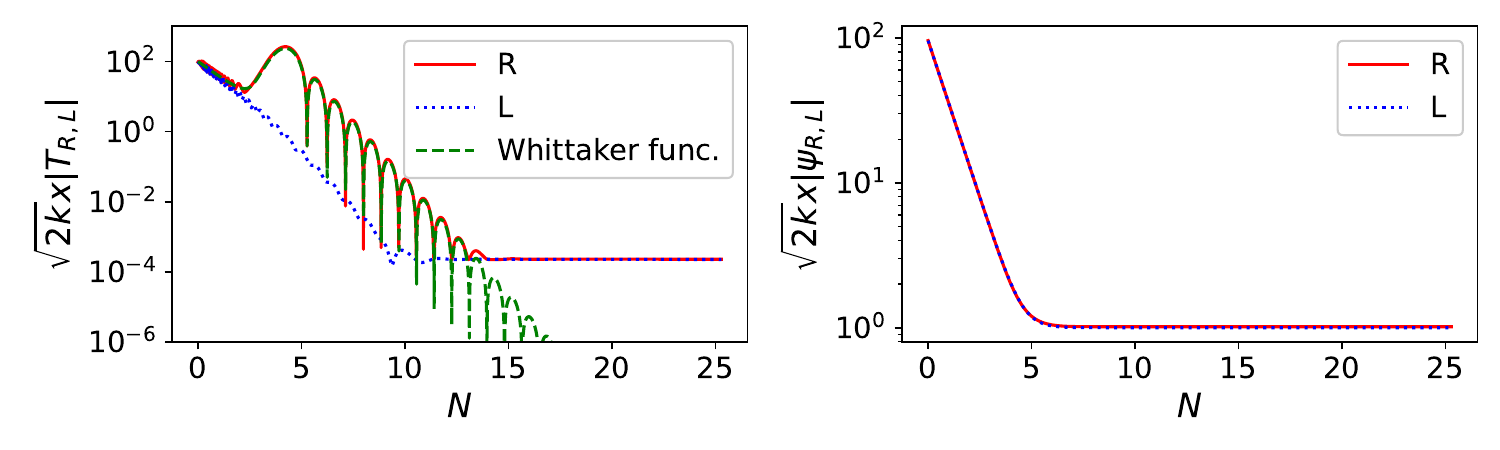}
\includegraphics[width=.95\textwidth]{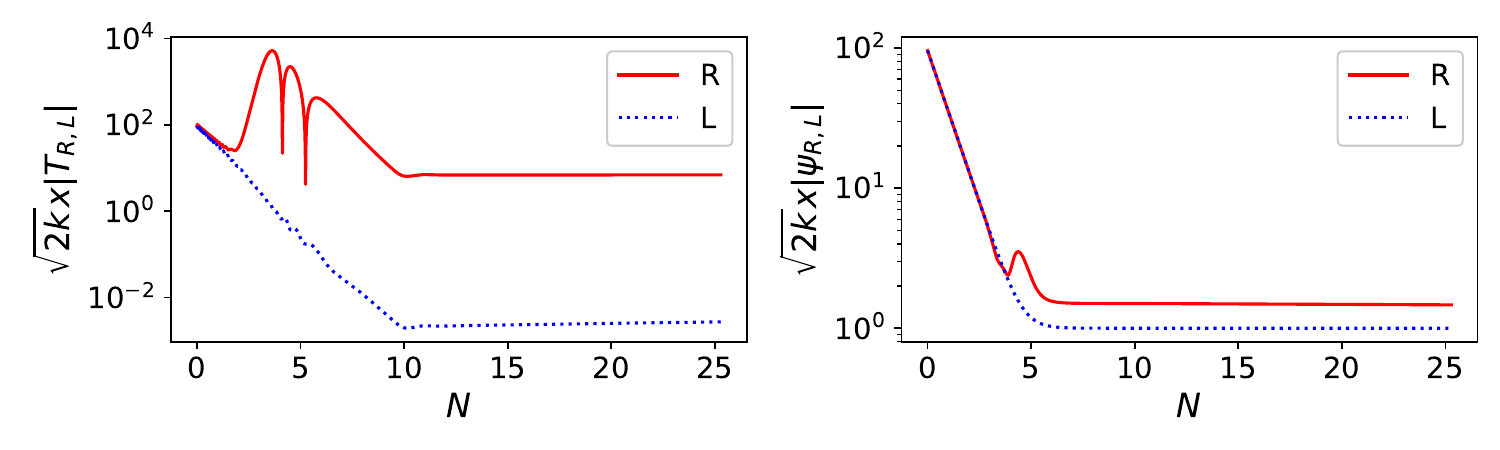}
\end{center}\caption[]{
$\sqrt{2k}x|T_{R,L}|$ and $\sqrt{2k}x|\psi_{R,L}|$ vs $N$ for $k=10^{-4}$ for runs~$\mu1$ with $m_Q=2.44$ (upper panel), $\mu$4 with  $m_Q=3.58$ (lower panel).
The solid red and dotted blue curves represent the numerical results for the right and left-handed polarization respectively. 
The dashed green curve shows the Whittaker solution for $T_R$ given by \Eq{Whittakersol}.
}\label{pTRL_8panels}\end{figure*}

As a test of our numerical implementation, we choose the initial value of $\mu$ and $g$ for one of the runs where the backreaction of the perturbations on the background evolution is negligible.
Run~$\mu$1 with $m_Q=2.44$ is an example of such a case. 
It is evident from \Fig{Q_and_chi_plot} that $Q$ remains constant
in time for run $\mu$1, as expected from the analytical results.
Furthermore, we have compared our numerical results for
$\sqrt{2k}(x|T_{R,L}|)$ and $\sqrt{2k}(x|\psi_{R,L}|)$ with the analytical solution
and show the comparison in the upper panel
of \Fig{pTRL_8panels} for $k=10^{-4}$.
Here, $x=-k\eta$ is the dimensionless time variable. 
In this figure, the solid red and dotted blue curves show the numerical result for
the right- and left-handed polarizations, respectively, and the dashed green curve
shows the analytic solution obtained using the homogeneous solution of $T_R$ in \Eq{treqn}.
This solution is given by \cite{Adshead:2013qp, Adshead:2013nka},
\begin{equation}\label{Whittakersol}
    T_R=\frac{1}{\sqrt{2k}}i^\beta W_{\beta,\alpha}(2ik\eta).
\end{equation}
Here, $W_{\beta,\alpha}(2ik\eta)$ is the Whittaker function with $\beta=-i (m_Q+\xi)$ and $\alpha=-i \sqrt{2 m_Q \xi-1/4}$.
The Whittaker function provides a good analytical approximation
for $T_R$ for a particular wave number approximately until the Hubble horizon crossing.
However, the solution starts to differ in the deep superhorizon regime due to the contribution from metric tensor perturbations.
It is evident from the upper panel of \Fig{pTRL_8panels} that our numerical solution matches well with the analytical solution in the regime where it is valid.

In the lower panel of \Fig{pTRL_8panels} we show the evolution of $\sqrt{2k}(x|T_{R,L}|)$ and $\sqrt{2k}(x|\psi_{R,L}|)$ for the case when backreaction is important.
We see that the gauge field tensor perturbations as well as the metric tensor perturbations are getting amplified for larger values of $m_Q$.
However, as we have seen in \Fig{Q_and_chi_plot}, such amplification drastically changes the behavior
of the background dynamics\footnote{Ref.~\cite{Ishiwata:2021yne} also found a deviation from the usual Chromo-Natural attractor, albeit not as drastic as the one described in the current work.
This can be attributed to the regime of couplings --and therefore the strength of the backreaction-- considered in ref.~\cite{Ishiwata:2021yne}.}. 

\begin{figure}
\begin{center}
\includegraphics[width=\columnwidth]{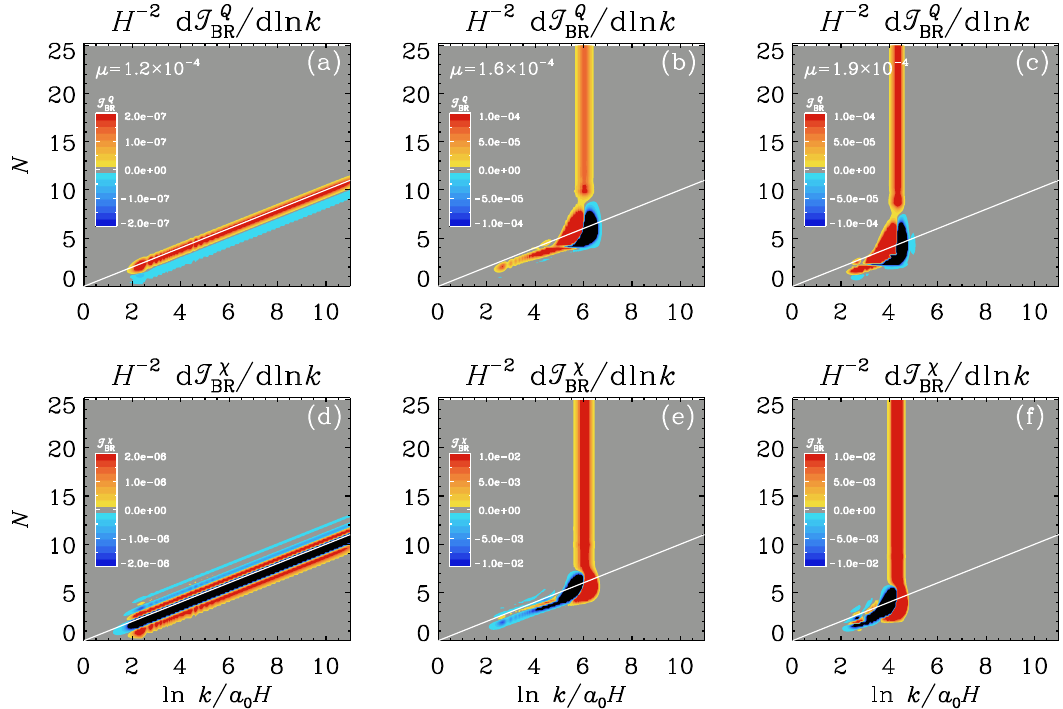}
\end{center}\caption[]{
The integrands of ${\cal T}_{\rm BR}^Q$ and ${\cal T}_{\rm BR}^\chi$, denoted
here by $d{\cal T}_{\rm BR}^Q/d\ln k$ and $d{\cal T}_{\rm BR}^\chi/d\ln k$, respectively, for
$\mu=1.2\times10^{-4}$ [run~$\mu$1 with $m_Q=2.44$, panels (a) and (d), no backreaction],
$\mu=1.6\times10^{-4}$ [run~$\mu$4 with $m_Q= 3.58$, panels (b) and (e)], and
$\mu=1.8\times10^{-4}$ [run~$\mu$5 with $m_Q= 4.19$, panels (c) and (f)].
The white line indicates the position of the comoving horizon.
}\label{pcomp_sample_mu1_mu4_mu5}\end{figure}

It is worth discussing the evolution of the backreaction integrals ${\cal T}^Q_{\rm BR}$ and ${\cal T}^{\chi}_{\rm BR}$ defined in equations \eqref{qbackint} and \eqref{chibackint}.
We show it in \Fig{pcomp_sample_mu1_mu4_mu5}.
From \Fig{pcomp_sample_mu1_mu4_mu5}, we conclude that most of the contribution
to the backreaction comes from a fixed narrow range of wave numbers.
This range is different for different values of $m_Q$.
For run~$\mu$4 with $m_Q=3.58$, the range is around $\ln k/(a_0 H)\approx 6$; see \Figsp{pcomp_sample_mu1_mu4_mu5}{b}{e}, while
for run~$\mu$5 with $m_Q=4.15$, it is $\ln k/(a_0 H)\approx 4$; see \Figsp{pcomp_sample_mu1_mu4_mu5}{c}{f}.
For run~$\mu$1 with $m_Q=2.44$, there is no backreaction; see \Figsp{pcomp_sample_mu1_mu4_mu5}{a}{d}.

We also performed variable $k$ range integrations for the same runs by calculating the backreaction integrals for modes around the comoving horizon. 
However, such a comoving integration scheme
leads to unphysical oscillatory features in the
background evolution, as we show
in \App{Artifacts}.

\subsection{Semi-analytical modelling}
\label{sec:semianalytical}
\begin{figure}
\centering
     \includegraphics[width=0.49\textwidth]{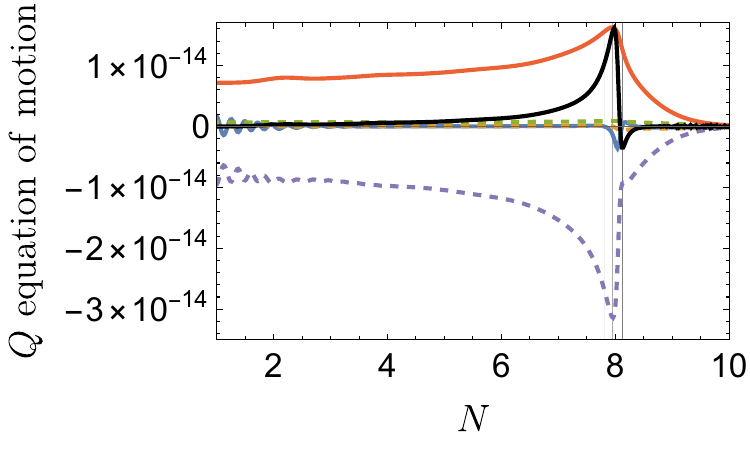}
    \includegraphics[width=0.49\textwidth]{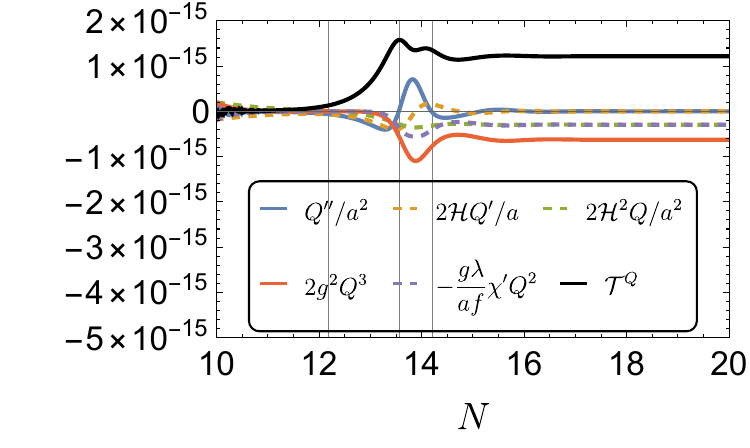}\\
    \includegraphics[width=0.49\textwidth]{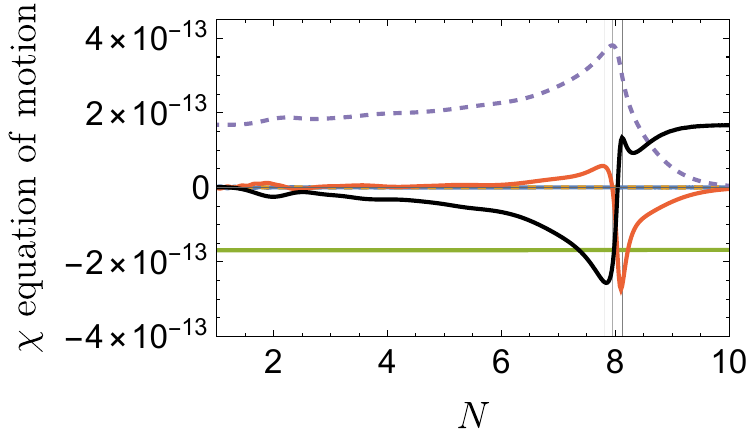}
   \includegraphics[width=0.49\textwidth]{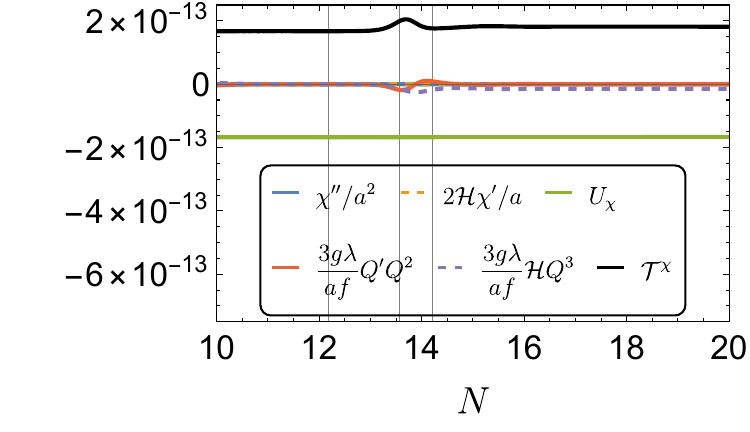}
\caption{\textit{Top left}: Contributions to the equation of motion for
the VEV of the gauge field, $Q$, with respect to the
number of $e$-folds at the initial stage and when backreaction occurs for the run $\mu$3 with $m_Q=3.29$. 
The three vertical gray grid lines correspond to the moments when $Q''=0$ , $Q'=0$, and then again $Q''=0$ (from left to right) respectively.
\textit{Top right}: Contributions to the equation of motion for $Q$ during the transition to the final attractor solution for the same run.
Vertical gray grid lines correspond to the moments when $Q=0$, $Q''=0$, $Q''=0$ (from left to right) respectively.
\textit{Bottom left}: Contributions to the equation of motion for the axion field, $\chi$, with respect to the number of $e$-folds at the initial stage and when backreaction is turned on for the same run as the top panels.
Grid lines are the same as in the top left panel.
\textit{Bottom right}: Contributions to the equation of motion for $\chi$ during the transition to the final attractor solution for the same run. Grid lines are the same as in the top right panel. }
\label{Fig:eombalance}
\end{figure}

In this section, we provide a semi-analytical analysis to approach
the new dynamical attractor solution.
In order to provide some intuition on the dynamics in the backreaction regime,
it is instructive to investigate the evolution of each contribution to
the equations of motion, \eqref{qeqn} and \eqref{chieqn}.
The time dependence of contributions is shown in \Fig{Fig:eombalance}.
Following the evolution of the backreaction terms $ {\cal T}^Q_{\rm
BR}$, $ {\cal T}^{\chi}_{\rm BR}$, three distinct phases of dynamics
can be distinguished.
From the top left and bottom left panels of \Fig{Fig:eombalance} one
can see that the backreaction contributions (solid black curves) grow
exponentially in absolute value up to around 8 $e$-folds.
We refer to the stage of exponential growth of backreaction
as Stage~I.
This leads to the destabilization of the ``standard'' chromo-natural attractor and pushes the system toward a new regime.
When the backreaction contributions become comparable to one of the
terms in the equations of motion, Stage~II begins.
At this stage, the backreaction terms change their behavior, start to
decrease, and eventually cross zero.
Following this stage, the system converges to the final solution (see the top
right and bottom right panels of \Fig{Fig:eombalance}) which we refer
to as Stage~III\footnote{During this stage the backreaction
terms dominate, questioning the validity of the linearized equations.
A full non-Abelian simulation would be required to fully explore  the late-time behavior, which is  outside the scope of the present work. We use the linear equations as an approximation to the full system.}.
The dynamics during different stages is described in more detail in
\App{appendix_Stages}.

\begin{figure*}\begin{center}
\includegraphics[scale=0.6]{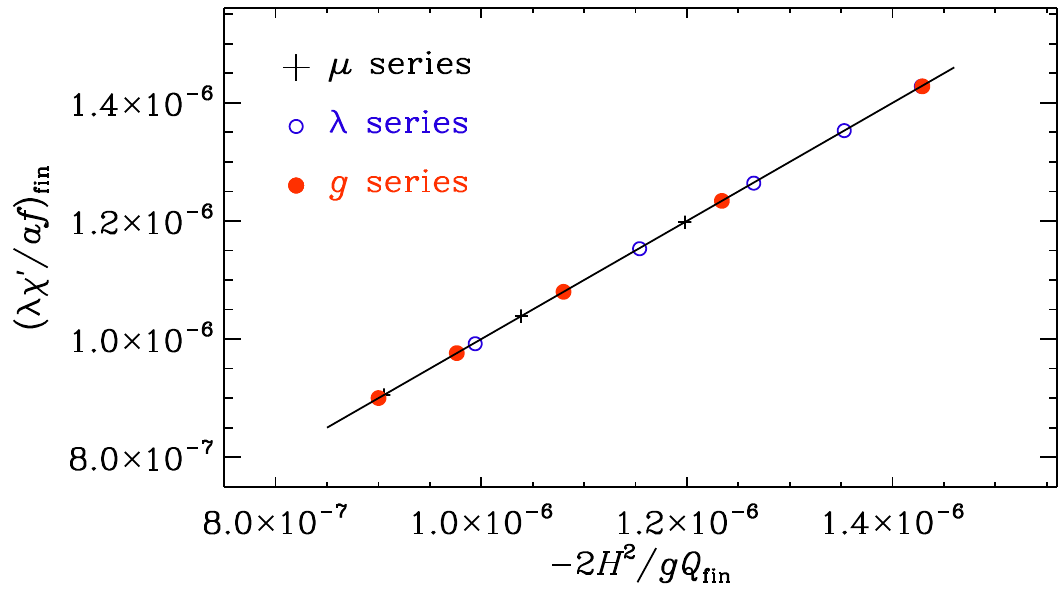}
\end{center}\caption[]{Verification of equation~\eqref{eqstage30}, showing
$(\lambda\chi'/af)_\mathrm{ fin}$ versus $-2H^2/gQ_\mathrm{fin}$, where $Q_\mathrm{fin}$ denotes the value of $Q$ at the final attractor for multiple series of runs from table~\ref{Tsummary}.
}\label{pchi_prime}\end{figure*}

Let us turn right away to the discussion of the new late-time dynamical attractor.
In Stage~III, all the contributions to the equations of motion become nearly constant.
In addition, the term $-(g\lambda/f)\,a\chi'Q^2$ becomes nearly
equal to the contribution $2 \mathcal{H}^2 Q$ in the equation
of motion \eqref{qeqn}\footnote{We used $\mathcal{H}'=a^2
\dot{H}+\mathcal{H}^2$ and $\dot{H}=0$ in  numerical simulations.},
which leads to
\begin{equation}
\begin{split}
 \frac{\lambda}{a f}\chi'\simeq -\frac{2H^2}{gQ},\label{eqstage30} \qquad
  Q\simeq {\rm const}.
\end{split}
\end{equation}
This solution resembles the original chromo-natural attractor solution given in \Eq{attractorCNI} with $Q=\rm {const}$ and just with the second term present that has an opposite sign.
It follows the late-time attractor $\xi\simeq -1/m_Q$.
In \Fig{pchi_prime}, we show that \Eq{eqstage30} does indeed hold.
With \eqref{eqstage30}, the equations of motion in Stage III become
\begin{gather}
4H^2Q+2g^2Q^3+{\cal T}^Q_{\rm BR}\simeq 0,\qquad
U_{\chi}+\frac{3g\lambda}{f}H Q^3+{\cal T}^{\chi}_{\rm BR}\simeq 0,\label{eqstage32}
\end{gather}
where we have taken into account that terms with derivatives in the last stage are negligible.

\begin{figure}
\centering
    \includegraphics[scale=0.6]{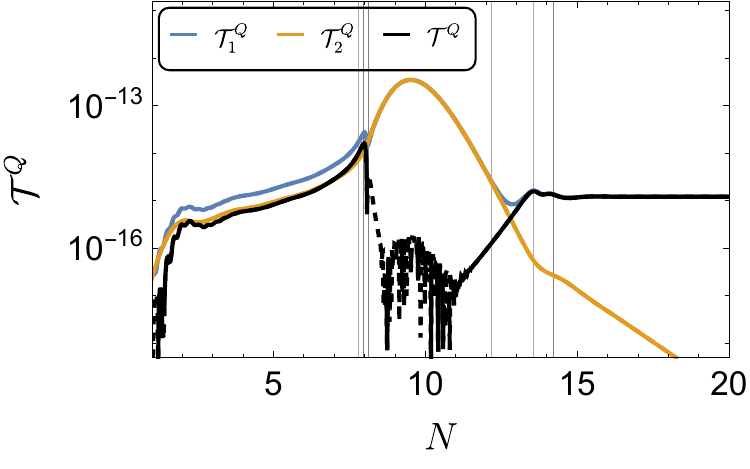}
    \includegraphics[scale=0.6]{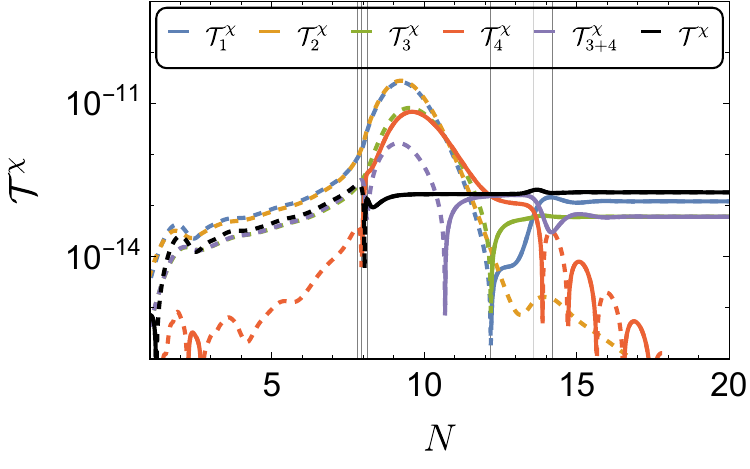}
\caption{\textit{Left}: Backreaction terms $ {\cal T}^Q_{\rm BR}$ and their contributions with respect to the number of $e$-folds  for the run $\mu$3 with $m_Q=3.29$. The dashed curve shows when $ {\cal T}^Q_{\rm BR}$ is negative.
\textit{Right}: Backreaction terms  ${\cal T}^{\chi}_{\rm BR}$ with contributions with respect to the number of $e$-folds for the same run. Grid lines are the same as in \Fig{Fig:eombalance}. Dashed curves represent negative values and solid curves show when functions are positive. } 
\label{Fig:backreaction}
\end{figure}

Let us now take a closer look into the time dependence of each component of the backreaction integrals \eqref{qbackint} and \eqref{chibackint}.
Backreaction terms may be written in conformal time as
\begin{gather}
 {\cal T}^Q_{\rm BR}=\frac{g}{3a^2}\left(  \xi H \langle|T|^2\rangle -\frac{1}{a}\langle k|T|^2\rangle\right)\equiv{\cal T}^Q_{1}+{\cal T}^Q_{2}, \label{backTermsQ} \\
 {\cal T}^{\chi}_{\rm BR}=-\frac{\lambda}{2 a^3 f} 
\left[ m_Q H \langle |T|^2\rangle'-\frac{1}{a}\langle k|T|^2\rangle'+
 \left(a m_Q H^2 + g Q'\right) \langle|T|^2\rangle 
\right]=\label{TchiAbs}\\
\equiv{\cal T}^{\chi}_{1}+{\cal T}^{\chi}_{2}+{\cal T}^{\chi}_{3}+{\cal T}^{\chi}_{4}, \label{backTermschi}
\end{gather}
where we have denoted the integrals over wave numbers by
\begin{gather}
\langle|T|^2\rangle=\int \frac{d^3 k}{(2\pi)^3}\,|T_R|^2,\quad
\langle k|T|^2\rangle=\int \frac{d^3 k}{(2\pi)^3}\,k\,|T_R|^2.
\end{gather}
We have checked that in all our cases, the contributions from $|T_L|$ to the backreaction are negligible.
In \Fig{Fig:backreaction}, we show the time evolution of the
different contributions to the ${\cal T}^Q_{\rm BR}$ and
${\cal T}^{\chi}_{\rm BR}$ integrals.
There is a clear correspondence between the background dynamics and
backreaction integrals, and vice versa.
We refer the interested reader to the detailed discussion in \App{appendix_Stages}.

\begin{figure*}[!ht]
\begin{center}
\includegraphics[scale=0.6]{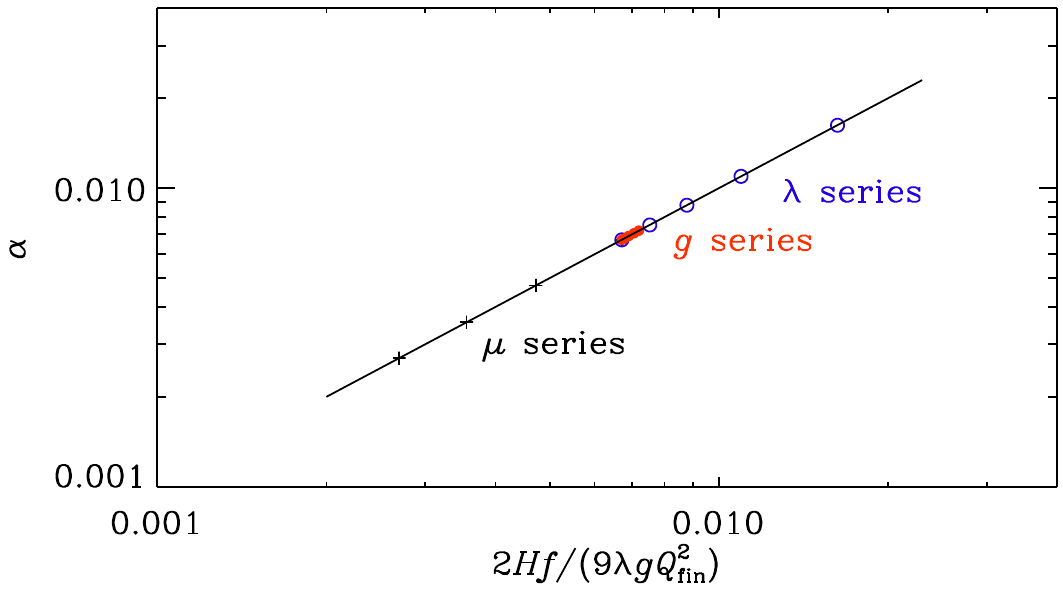}
\end{center}\caption[]{Dependence of the parameter 
$\alpha$ on $Q_{\rm fin}$ for three series of runs.
}\label{palp_versus_Q}
\end{figure*}

The backreaction integrals include crucial information that governs the evolution of the whole system.
It is therefore convenient to introduce a new parameter that quantifies the ratio of the two backreaction integrals
\begin{equation}
    \frac{{\cal T}^{Q}_{\rm BR}}{{\cal T}^{\chi}_{\rm BR}}\equiv \alpha.
\end{equation}
From equations~\eqref{eqstage32}, one can find the relation between $\chi$ and $Q$ on the final attractor
\begin{equation}
  U_{\chi}= - \frac{3g\lambda}{f}H Q^3+\frac{1}{\alpha}\left(4 H^2 Q+2g^2 Q^3 \right).\label{UQattractor}
\end{equation}
At the final stage it holds
${\cal T}^Q_{\rm BR}\simeq {\cal T}^Q_{1}$ and ${\cal T}^{\chi}_{\rm BR}\simeq {\cal T}^{\chi}_{1}+{\cal T}^{\chi}_{3}$,
so the parameter $\alpha$ can be expressed as
\begin{equation}
    \alpha=\frac{2}{9}\frac{Hf}{\lambda g \, Q_\mathrm{fin}^2}, \label{alpha}
\end{equation}
where we used $\langle |T|^2\rangle' \simeq 2 a H \langle|T|^2\rangle$
and $Q_\mathrm{fin}$ denotes the value of $Q$ on the final attractor.
The dependence of $\alpha$ on the parameters $\lambda$, $g$, and $\mu$ is confirmed in \Fig{palp_versus_Q} for runs of the three families.
To sum up, the late-time dynamical attractor is given by equations \eqref{eqstage30}, \eqref{UQattractor} and \eqref{alpha}.

\subsection{Observational signatures}
\label{ObservationalSignatures}
The GW energy density power spectrum can be approximated as \cite{Caprini:2018mtu}
\begin{equation}
\Omega_{\rm GW}(k)=\frac{3}{128}\Omega_{\rm rad}{\cal P}_h^{\rm tot}(k)\left[ \frac{1}{2}\left(\frac{k_{\rm eq}}{k}\right)^2+\frac{16}{9}\right],
\end{equation}
where $\Omega_{\rm rad}\simeq h^{-2}\,2.47 \times 10^{-5}$ is the present radiation density parameter and
$k_{\rm eq}\simeq 1.3 \times 10^{-2} \, {\rm Mpc}^{-1}$ is the wave number entering the horizon at matter-radiation equality.
Here, $h$ is defined such that $H_0=100\, h \, {\rm km} \, {\rm s}^{-1} \, {\rm Mpc}^{-1}$ is the Hubble parameter at the present epoch.
To express $\Omega_{\rm GW}$ as a function of frequency, we use $f\simeq 1.5 \times 10^{-15} \, (k/{\rm Mpc}^{-1}) \, {\rm Hz}$.
Here, the frequency is not to be confused with the axion decay constant, which is also called $f$.

\begin{figure*}
\begin{center}
\includegraphics[scale=0.6]{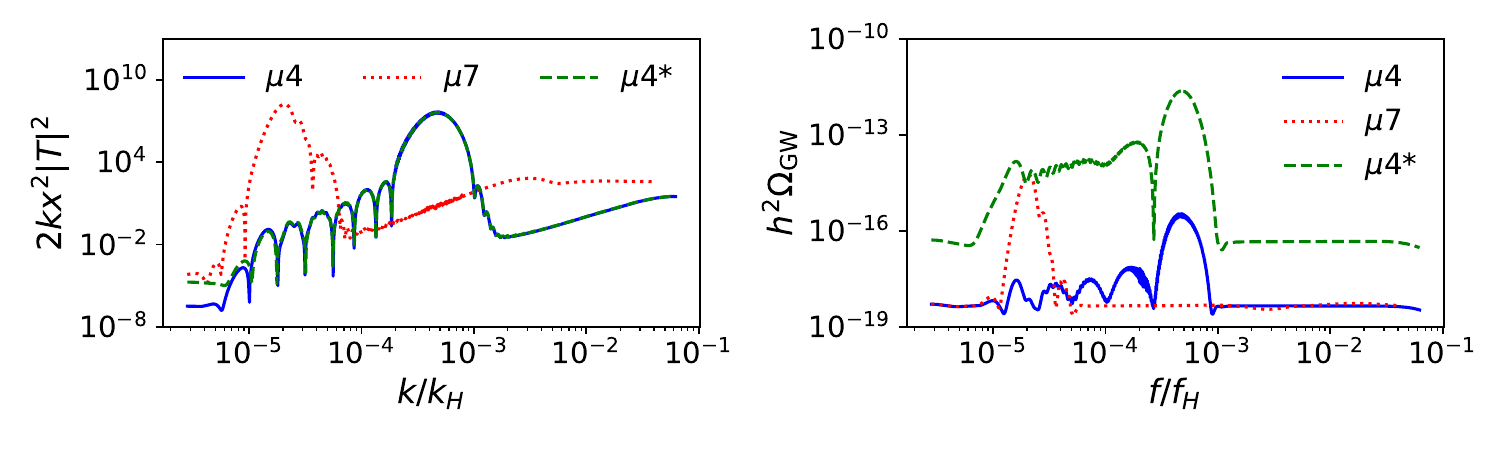}
\end{center}\caption[]{$2k x^2 |T|^2$ vs $k/k_H$ and $h^2\Omega_{\rm GW}$ vs $f/f_H$ is shown for
the runs $\mu$4, $\mu4^*$, and $\mu$7.
}
\label{omegagwplot}
\end{figure*}

In \Fig{omegagwplot}, we show $2k x^2 |T|^2$ vs $k/k_H$ and $\Omega_{\rm GW}$ vs $f/f_H$ for runs $\mu$4, $\mu$7, and $\mu4^*$  with $m_Q=3.58, \, 5.15, \, 3.78$ respectively (solid blue, dotted red, and dashed green curves).
The run denoted as $\mu$4$^*$ is analogous to the $\mu$4 run but with a Hubble parameter $H$ larger by a factor of 10.
In this figure, we have normalized the wave number and GW frequency
using the values corresponding to the Hubble horizon size ($k_H$) at
the end of inflation and the corresponding frequency ($f_H$).
These quantities are given by
\begin{equation}\label{kHfH}
k_H = \frac{a_e}{a_0} H= 2.3 \times 10^{22} \text{Mpc}^{-1} \frac{H}{1.04 \times 10^{-6}\,M_\text{pl}}, \quad
f_H = \frac{k_H}{2\pi} = 3.5 \times 10^{7} \text{Hz}.
\end{equation}
Here, $H = 1.04 \times 10^{-6}M_\text{pl}$ and we assume an adiabatic evolution of the Universe to calculate $a_e/a_0$, given by
\begin{equation}
\frac{a_e}{a_0} = \left(\frac{g_{0s}}{g_{rs}}\right)^{1/3}\frac{T_0}{T_r} = 5.8 \times 10^{-29} \frac{g_{0s}}{3.94} \frac{106.75}{g_{rs}} \frac{T_0}{2.73 \text{K}} \frac{1.3 \times 10^{15}}{T_r}.
\end{equation}
In the above expression, $g_{\ast s}$ and $g_{0s}$ denote the effective
degrees of freedom in the entropy at the end of inflation and the present
epoch, respectively, and $T_r$ denotes the reheating temperature assuming
instantaneous reheating.
We estimate $T_r$ by using the relation
$3H^2M_\text{pl}^2 = (\pi^2/30)g_rT_r^4$. It is worth noting that we normalize the frequency in \Fig{omegagwplot} by assuming that the end of the simulation is also the end of inflation. However, if the end of the simulation is $N_e$ e-folds before the end of inflation, the wave number and corresponding frequency must be multiplied by $e^{-N_e}$.

As is evident from \Fig{omegagwplot}, the modes that are amplified around horizon crossing give the largest contribution to $\Omega_{\rm GW}$. 
The oscillations observed in $\Omega_{\rm GW}$ are related to the oscillations in $T_R$, as illustrated in the left panel of \Fig{omegagwplot}.
The higher initial value of $m_Q$ in run~$\mu$7 results in a higher peak value of $\Omega_{\rm GW}$.
As demonstrated in \Fig{omegagwplot}, the higher value of $H$ leads to a higher value of $\Omega_{\rm GW}$.

\begin{figure*}
\begin{center}
\includegraphics[scale=0.55]{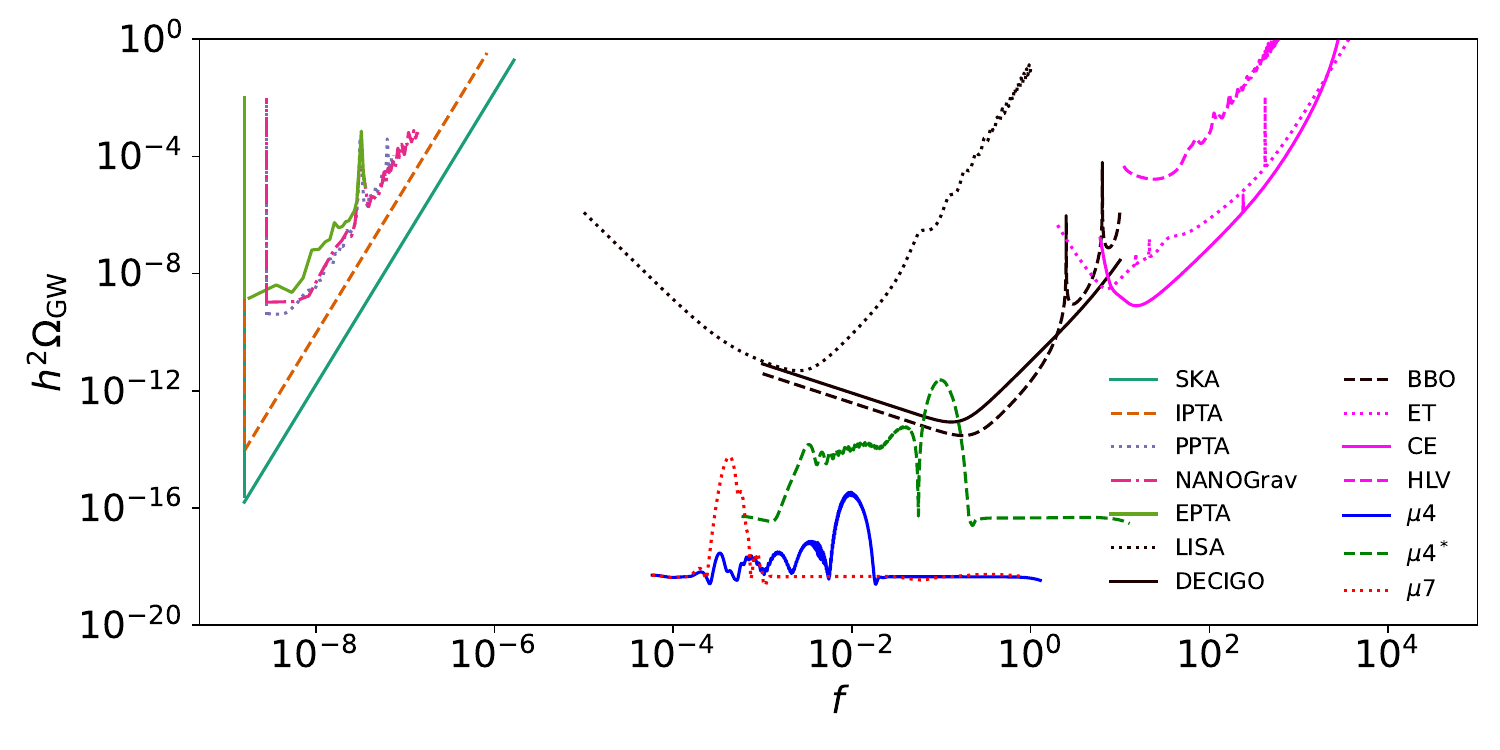}
\end{center}\caption[]{$h^2\Omega_{\rm GW}$ vs $f$ along with the sensitivity curves of various detectors \cite{Schmitz:2020syl}.
}\label{omegagwplot_with_detect}
\end{figure*}

In \Fig{omegagwplot_with_detect}, we show $h^2 \Omega_{\rm GW}$ for runs~$\mu$4, $\mu$7, and $\mu4^*$ ($m_Q=3.58$, $5.15$, and $3.78$, respectively),
along with sensitivity curves for various GW detectors.
The sensitivity curves are obtained by using the ``strain noise power spectra'' file available in the Zenodo online repository \cite{schmitz_2020_3689582} associated with ref.~\cite{Schmitz:2020syl}.
Our simulation covers approximately 25 $e$-folds. To provide a comparison of the obtained $h^2 \Omega_{\rm GW}$ with the sensitivity of upcoming GW detectors DECIGO and BBO, we rescale the GW frequency such that the end of our simulation occurs 14 $e$-folds before the end of inflation. For run~$\mu4^*$, the peak in $\Omega_{\rm GW}$ at $f=0.1\,\text{Hz}$ is potentially detectable.

\section{Summary and discussion }\label{sec:summary}

In this work, we simulated an axion-SU(2) sector, which is a spectator
during inflation, meaning that its energy density is subdominant to the
inflaton and that both the Hubble rate and the density perturbations
are unaffected by its presence.
In our simulations, the fluctuations are computed using the linearized
equations of motion. Their effect on the axion and gauge field
VEV are computed self-consistently, but averaged over the simulation domain.
This method is similar in spirit to the one followed in
ref.~\cite{Garcia-Bellido:2023ser} in the case of axion-$U(1)$ inflation,
where the gauge fields are computed using linear equations of motion and
their collective effect is considered a background quantity and added to
the corresponding background equations.
In the case of the Abelian model, an oscillatory result was found that
can be understood in the following way: an increase in the axion velocity
leads to an increase in the gauge field amplification.
This results in an increased backreaction on the rolling axion through
the $\langle \EE \cdot \BB\rangle$ term; see \App{ASsolution}.
This backreaction leads to a slow-down of the axion, which, in turn,
reduces the subsequent amplification of gauge fields.
Since the gauge fields  red-shift after their production, their
backreaction will also reduce, leading to a speed-up of the axion and
the whole process will start anew, thus leading to periodic bursts of
gauge field production.

In the non-Abelian case, the initial stage is similar to the Abelian
case: as the axion picks up speed, the gauge fields (a tensor mode in
this case) are exponentially amplified.
This leads to a backreaction on the equations of motion that define
the VEV of both the axion field as well as the SU(2) sector.
This leads to both a slow-down of the axion, as well as a completely
new sign-flipped value for the gauge field VEV.
Furthermore, in this new regime, the superhorizon tensor modes of
the gauge field evolve as $T_{R,L}\propto 1/\eta$.
This leads to the absence of the periodic behavior found in the Abelian
case, because the backreaction of the gauge field fluctuations onto the
background quantities does not diminish with time (to  lowest order
in slow-roll).

Having revealed this new regime, several questions remain to be answered.
An intriguing relation was found between the backreaction terms in
the axion and gauge VEV, leading to a universal relation between the
parameters of the potential and the late-time value of the gauge VEV.
However, we are not able to predict the gauge VEV itself in this new attractor.
We believe that the initial value of the gauge VEV (in the original
chromo-natural attractor) plays a role in determining its late-time value.

Furthermore, our analysis neglects spatially dependent backreaction
effects that can lead to mode-mode coupling of the gauge field, as well
as the excitation of scalar fluctuations in the axion sector.
It has been shown in ref.~\cite{Figueroa:2023oxc} that space-dependent
backreaction effects can be very important in the Abelian case.
More importantly, large values of the gauge fields may lead to 
 non-Abelian interactions becoming important and invalidate the linear approximation. 
While we expect the initial growth of the fluctuations and destabilization of the ``standard'' chromo-natural solution to hold, providing conclusive proof on the existence and stability of the late-time attractor requires solving the full system
on a lattice, without making any linear or Hartree-type approximations.
In addition, non-Abelian fields coupled to a rolling axion have been
argued to lead to warm inflation \cite{DeRocco:2021rzv, Berghaus:2019whh}.
It is an intriguing possibility to examine if a system can transition from the chromo-natural inflation attractor to warm inflation, either in the original chromo-natural inflation formulation or in spectator models.

Furthermore, our calculation was performed with a constant Hubble scale,
in an exact de-Sitter background.
While this can be an excellent approximation for several inflationary
models, it does not allow us to probe the evolution of this new attractor
close to the end of inflation, when $|\dot H / H^2| \sim 1$.

Finally, the flipped sign of the gauge field VEV provides the possibility
of amplifying the subdominant helicity of gauge tensor modes.
Further analysis of this is left for future work, as it can provide
interesting scale-dependent observables.

\section*{Acknowledgments}

We thank Peter Adshead, Tomohiro Fujita, Azadeh Maleknejad, and Lorenzo Sorbo for useful comments on the draft, and
Alberto Roper Pol for information regarding the sensitivity curves of various GW detectors.
The work of O.I.\ was supported by the European Union's Horizon 2020
research and innovation program under the Marie Skłodowska-Curie grant
agreement No.~101106874.
E.I.S. is supported through the ``la Caixa'' Foundation (ID 100010434) and the European Union's Horizon 2020 research
and innovation programme under the Marie Sk\l{}odowska-Curie grant agreement No 847648 under the fellowship code
LCF/BQ/PI20/11760021.
R.S.\ was supported through grant No.~ERC HERO-810451 from the European Research Council.
A.B.\ acknowledges the grants No.~2019-04234 from the Swedish Research Council (Vetenskapsr{\aa}det)
and the ASA ATP grant No.~80NSSC22K0825.
O.I.\ is grateful to the University of Leiden for hospitality, where parts of this work have been completed.
We thank the Swedish National Allocations Committee for providing computing resources at the Center for Parallel Computers
at the Royal Institute of Technology in Stockholm and the National Supercomputer Centre (NSC) at Link\"oping.
Nordita was sponsored by Nordforsk.

\paragraph{Data availability.}
The source code used for the numerical solutions of this study, the
{\sc Pencil Code}, along with the module \texttt{special/axionSU2back.f90}
(after October 6, 2023), used in the present study, are freely available~\cite{JOSS}.
The numerical data and input files are available on Zenodo; see ref.~\cite{DATA}.

\appendix
\section{Full set of parameters}
\label{SetOfParameters}
The full set of parameters used in the simulations is shown in Table~\ref{Tsummary}.
The fiducial parameters are $\mu=1.5\times10^{-4}$, $g=1.11\times10^{-2}$, $f=0.003$
and $\lambda=500$.
The runs are grouped separately for
runs~$\mu1$--$\mu8$ with $1.2\leq\mu/10^{-4}\leq2.45$,
runs~$g1$--$g5$ with $1.11\leq g/10^{-2}\leq3$, and
runs~$\lambda1$--$\lambda6$ with $100\leq\lambda\leq600$.
Only the runs with negative values of $Q_{\rm fin}$ have
undergone backreaction.
For run~$\mu4^*$, we have used $H=1.04\times10^{-5}\,M_\text{pl}$,
which leads to a higher value of $\Omega_{\rm GW}$; see \Sec{ObservationalSignatures}.

\begin{table}
\begin{center}
\begin{tabular}{ccccccrr}
\hline
Run & $\mu$ & $g$ & $\lambda$ & $m_{Q_0}$ & $Q_0$ & $Q_{\rm fin}\qquad$ & $\alpha\qquad\;$ \\
\hline
$\mu$1&$1.20\times10^{-4}$&$1.11\times10^{-2}$&500&$2.44$&$2.29\times10^{-4}$&$2.26\times10^{-4}$&$-1.67\times10^{-2}$\\
$\mu$2&$1.40\times10^{-4}$&$1.11\times10^{-2}$&500&$3.00$&$2.81\times10^{-4}$&$2.77\times10^{-4}$&$-1.59\times10^{-2}$\\
$\mu$3&$1.50\times10^{-4}$&$1.11\times10^{-2}$&500&$3.29$&$3.08\times10^{-4}$&$-1.36\times10^{-4}$&$6.71\times10^{-3}$\\
$\mu$4&$1.60\times10^{-4}$&$1.11\times10^{-2}$&500&$3.58$&$3.36\times10^{-4}$&$-1.45\times10^{-4}$&$5.91\times10^{-3}$\\
$\,\;\mu4^*$&$1.60\times10^{-3}$&$1.11\times10^{-2}$&500&$3.78$&$3.40\times10^{-3}$&$-1.46\times10^{-3}$&$5.90\times10^{-4}$\\
$\mu$5&$1.80\times10^{-4}$&$1.11\times10^{-2}$&500&$4.19$&$3.93\times10^{-4}$&$-1.63\times10^{-4}$&$4.72\times10^{-3}$\\
$\mu$6&$1.90\times10^{-4}$&$1.11\times10^{-2}$&500&$4.51$&$4.22\times10^{-4}$&$-1.71\times10^{-4}$&$4.27\times10^{-3}$\\
$\mu$7&$2.10\times10^{-4}$&$1.11\times10^{-2}$&500&$5.15$&$4.82\times10^{-4}$&$-1.88\times10^{-4}$&$3.55\times10^{-3}$\\
$\mu$8&$2.45\times10^{-4}$&$1.11\times10^{-2}$&500&$6.32$&$5.92\times10^{-4}$&$-2.15\times10^{-4}$&$2.70\times10^{-3}$\\
\hline
$g$1&$1.50\times10^{-4}$&$1.11\times10^{-2}$&500&$3.29$&$3.08\times10^{-4}$&$-1.36\times10^{-4}$&$6.71\times10^{-3}$\\
$g$2&$1.50\times10^{-4}$&$1.50\times10^{-2}$&500&$4.02$&$2.79\times10^{-4}$&$-1.17\times10^{-4}$&$6.77\times10^{-3}$\\
$g$3&$1.50\times10^{-4}$&$2.00\times10^{-2}$&500&$4.87$&$2.53\times10^{-4}$&$-1.00\times10^{-4}$&$6.91\times10^{-3}$\\
$g$4&$1.50\times10^{-4}$&$2.50\times10^{-2}$&500&$5.65$&$2.35\times10^{-4}$&$-8.86\times10^{-5}$&$7.06\times10^{-3}$\\
$g$5&$1.50\times10^{-4}$&$3.00\times10^{-2}$&500&$6.38$&$2.21\times10^{-4}$&$-8.01\times10^{-5}$&$7.20\times10^{-3}$\\
\hline
$\lambda$1&$1.50\times10^{-4}$&$1.11\times10^{-2}$&600&$3.09$&$2.90\times10^{-4}$&$2.87\times10^{-4}$&$-1.31\times10^{-2}$\\
$\lambda$2&$1.50\times10^{-4}$&$1.11\times10^{-2}$&500&$3.29$&$3.08\times10^{-4}$&$-1.36\times10^{-4}$&$6.71\times10^{-3}$\\
$\lambda$3&$1.50\times10^{-4}$&$1.11\times10^{-2}$&400&$3.54$&$3.32\times10^{-4}$&$-1.44\times10^{-4}$&$7.53\times10^{-3}$\\
$\lambda$4&$1.50\times10^{-4}$&$1.11\times10^{-2}$&300&$3.90$&$3.65\times10^{-4}$&$-1.54\times10^{-4}$&$8.77\times10^{-3}$\\
$\lambda$5&$1.50\times10^{-4}$&$1.11\times10^{-2}$&200&$4.46$&$4.18\times10^{-4}$&$-1.69\times10^{-4}$&$1.10\times10^{-2}$\\
$\lambda$6&$1.50\times10^{-4}$&$1.11\times10^{-2}$&100&$5.62$&$5.27\times10^{-4}$&$-1.96\times10^{-4}$&$1.62\times10^{-2}$\\
\hline
\end{tabular}
\caption{\label{Tsummary}
Summary of runs for the $g$, $\lambda$, and $\mu$ series.
For each series, the first line refers to the fiducial run with
$\mu=1.5\times10^{-4}$, $g=1.11\times10^{-2}$, $f=0.003$ and $\lambda=500$.
The asterisk on run~$\mu4^*$ indicates that here
$H=1.04\times10^{-5}\,M_\text{pl}$ is 10 times larger than usual. 
}\end{center}
\end{table}

\section{Artifacts from not resolving the superhorizon modes}
\label{Artifacts}

\begin{figure}
\begin{center}
\includegraphics[width=\columnwidth]{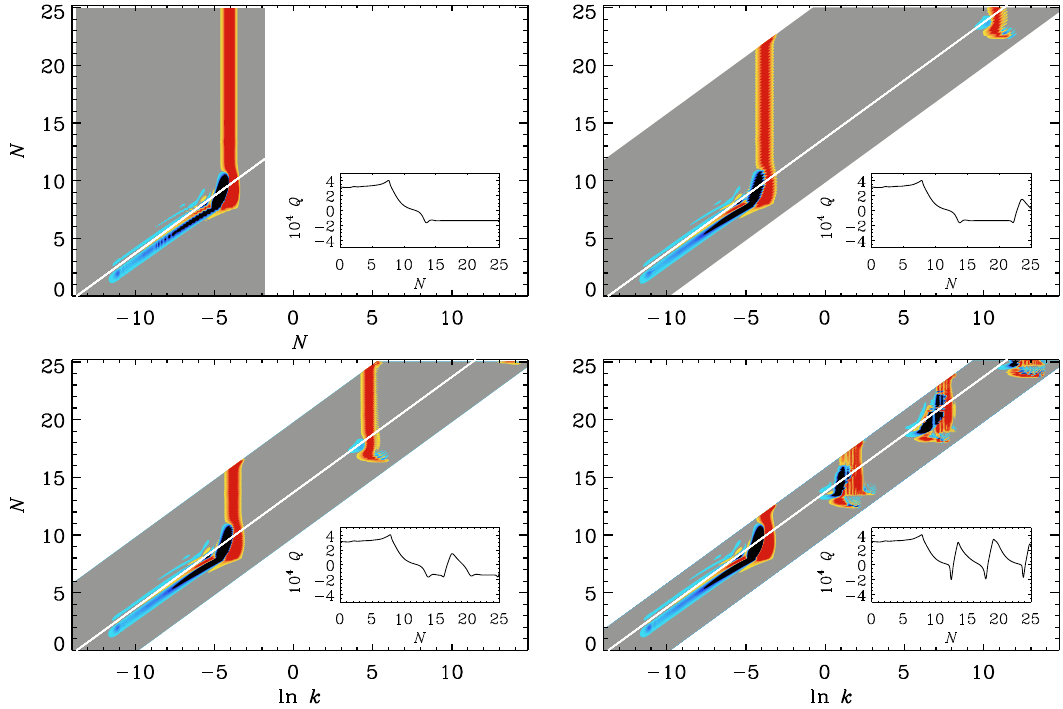}
\end{center}\caption[]{
${\cal T}_{\rm BR}^Q$ for $(Q_0,\mu)/10^{-4}=(3.2,\,1.5)$
without shift for $0\leq\ln k\leq12$ (first panel),
and with shift for $-12\leq\ln k/aH \leq4$,
$-12\leq\ln k/aH \leq4$, $-6\leq\ln k/aH \leq4$, and
$-2\leq\ln k/aH \leq4$ (the other 3 panels).
The insets show the corresponding evolution of $Q$.
}\label{pcomp_sample_3p2}\end{figure}

One might have expected that it is important to resolve the modes around
the comoving horizon.
Looking at \Fig{pcomp_sample_mu1_mu4_mu5}, this is not obvious, however.
Once backreaction becomes important, most of the contributions to the
backreaction come from a fixed band of wave numbers.
It is instructive to examine the results where we allow for
the possibility to move the range of integration to a comoving strip with
\begin{equation}
n_{\min}\leq\ln[k/a(\eta) H] \leq n_{\max}.
\end{equation}
The result of numerical simulation using a comoving strip of wave numbers is shown in \Fig{pcomp_sample_3p2}.

As we see from the insets of \Fig{pcomp_sample_3p2},
the corresponding evolution of $Q$ is different in cases
where the modes in the proximity of the comoving horizon
are resolved at the expense of not capturing any more the
strongly superhorizon modes.
We can conclude that this causes numerical artifacts that look like periodic bursts of gauge field production during inflation.

\section{Detailed description of backreaction stages }\label{appendix_Stages}
\begin{figure}
\centering
     \includegraphics[width=0.52\textwidth]{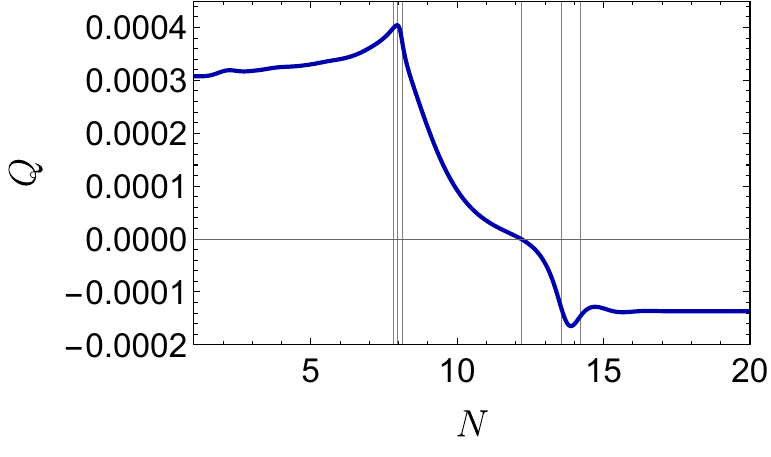}
    \includegraphics[width=0.47\textwidth]{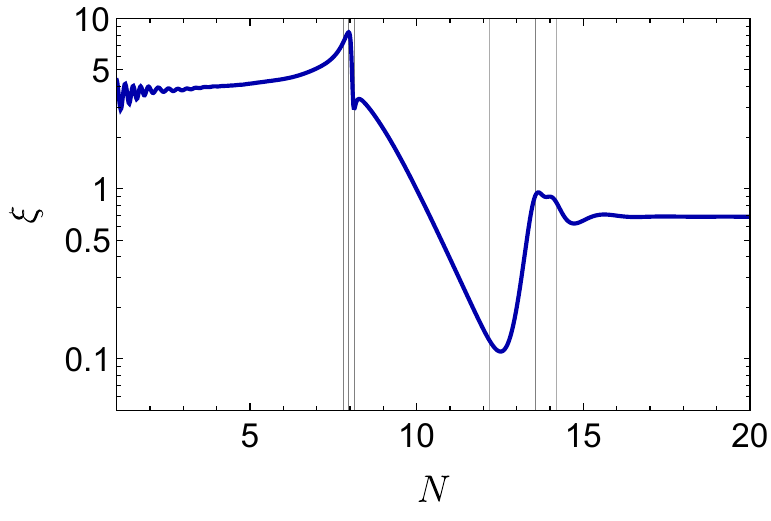}\\
\caption{\textit{Left}: VEV of the gauge field $Q$ versus the number of $e$-folds for run~$\mu$3 with $m_Q=3.29$.
Grid lines are the same as in figure~\ref{Fig:eombalance}.
\textit{Right}: Evolution of the $\xi$ parameter, defined in \Eq{mQxidef} with respect to the number of $e$-folds, for the same run as in the left panel.}
\label{Fig:Qxi}
\end{figure}

At the initial stage, referred to here as Stage~I, the solution follows
the chromo-natural attractor solution \eqref{attractorCNI}, where the
three terms $-(g\lambda/af) \chi' Q^2$, $2g^2 Q^3$, and
$2\mathcal{H}^2Q/a^2$ balance each other in \eqref{qeqn}.
Note that, in the present case of $H=\mathrm{const}$, we have
$\mathcal{H}'+\mathcal{H}^2=2\mathcal{H}^2$.
When the backreaction ${\cal T}^Q_{\rm BR}$ becomes important, the
contribution $-(g\lambda/af)\,\chi'Q^2$ becomes more dominant
compared to the rest of the terms (see purple dashed curve in the
top left and bottom left of figure \ref{Fig:eombalance}).
To compensate for the increase of the sum ${\cal T}^Q_{\rm BR}$ and
$-(g\lambda/af)\,\chi'Q^2$ terms, the $Q''$ contribution becomes negative.
It causes the change in the sign of $Q'$.
This changes the behavior and turns on the $Q'$ term,
${\cal T}^{\chi}_{4}$, in the integral ${\cal T}^{\chi}_{\rm BR}$
of \Eq{backTermschi}, which produces a bump in $\xi$; see \Fig{Fig:Qxi}.
This happens around $N=8$ $e$-folds.
The change in $\xi$ makes the two terms ${\cal T}^Q_{1}$ and
${\cal T}^Q_{2}$ in \eqref{backTermsQ} almost cancel each other, see \Fig{Fig:backreaction}.
As a result, the ${\cal T}^Q_{\rm BR}$ term becomes first negative and is then close to zero.
This causes a decrease of $Q$.
The steps at Stage~I can be described by the following chain sequence:
\begin{equation*}
 {\cal T}^Q_{\rm BR} \rightarrow   Q'' \rightarrow Q' \rightarrow {\cal T}^{\chi}_{\rm BR} \rightarrow \xi(\chi') \rightarrow {\cal T}^Q_{\rm BR} \rightarrow Q.
\end{equation*}

The next stage, Stage~II, is characterized by a continuous decrease of $Q$.
The backreaction ${\cal T}^Q_{\rm BR}\approx 0$ remains small, and ${\cal T}^{\chi}_{\rm BR}\approx \rm{const}$.

The last stage, Stage III, begins when the gauge field VEV reaches zero, i.e., $Q=0$. 
This changes the sign of terms with $m_Q$, i.e., ${\cal T}^{\chi}_1$ and ${\cal T}^{\chi}_3$ from \eqref{backTermschi}.
This governs the change in $\xi$ and causes the inflection of the ${\cal T}^{Q}_1$ contribution.
As a result, the solution arrives at the final attractor \eqref{eqstage30} with
${\cal T}^{\chi}={\cal T}^{\chi}_1+{\cal T}^{\chi}_3\approx \rm{const}$ and ${\cal T}^{Q}={\cal T}^{Q}_1\approx \rm{const}$.
At Stage~III, we observe the following chain sequence:
\begin{equation*}
Q \rightarrow m_Q \rightarrow {\cal T}^{\chi}_{\rm BR} \rightarrow \xi(\chi') \rightarrow  {\cal T}^Q_{\rm BR} \rightarrow   Q.
\end{equation*}
The three stages of evolution are summarized in Table~\ref{table1}.

\begin{table}
\begin{center}
\begin{tabular}{|p{4cm}|p{5cm}|p{5cm}| }
 \hline
 Stage I& Stage II  &Stage III\\
 \hline
 $ {\cal T}^Q_{\rm BR}\propto \exp\left({\cal O}(1)N\right)$  & $ {\cal T}^Q_{\rm BR}\approx  0$    & $ 4H^2Q+2g^2Q^3+{\cal T}^Q_{\rm BR}\simeq 0$ \\
 $|{\cal T}^{\chi}_{\rm BR}|\propto \exp\left({\cal O}(1) N\right)$& $U_{\chi}+(3g\lambda/f) H Q^3+{\cal T}^{\chi}_{\rm BR}\simeq 0$ &
 $U_{\chi}+(3g\lambda/f) H Q^3+{\cal T}^{\chi}_{\rm BR}\simeq 0$\\
 \hline
\end{tabular}
\caption{\label{table1}The three stages of dynamics in axion-SU(2) inflation with backreaction. }
\end{center}
\end{table}

\section{Dynamical attractor, gravitational waves and the Anber--Sorbo solution}
\label{ASsolution}

It is interesting to compare the dynamical attractor solution found in this work to the Anber--Sorbo (AS) solution \cite{Anber:2009ua}, which applies to axion-U(1) inflation.
We start by briefly reviewing the AS solution.
While it has recently been shown to be unstable \cite{vonEckardstein:2023gwk}, its analytical derivation can still be illuminating when compared to the dynamical attractor found in the present work.

We assume an exact de-Sitter expansion for simplicity, meaning that conformal time $\eta$ is  $\eta = -1/Ha = -e^{-Ht}/H$. The equation of motion for the background axion field is
\begin{equation}
\ddot\phi +3H\dot\phi +V_{,\phi} = {\alpha\over f} \left \langle \EE\cdot\BB \right \rangle  ,
\end{equation}
where the gauge fields follow the linear equations (in conformal time)
\begin{equation}
{d^2 A_{\pm}\over d\eta^2} + \left (
k^2 \mp {2 k \xi\over \eta}
\right ) A_{\pm}=0 \, ,
\end{equation}
and $\xi = \alpha \dot\phi / (2fH)$ is taken to be constant during inflation.
At $|\eta| = 2|\xi| /k$, one of the polarizations becomes tachyonic and for $|k\eta| \ll 2\xi$,
the solution of the amplified polarization $A_+$ is approximated as
\begin{equation}
A_+\simeq {1\over \sqrt{2k}} \left (
k\over 2\xi aH
\right )^{1/4} e^{\pi\xi} e^{-2\sqrt{2\xi k /aH}},
\end{equation}
where we take $\xi>0$. The backreaction is
\begin{equation}
\left \langle \EE\cdot\BB \right \rangle = - {1\over a^4} \int {d^3k\over (2\pi)^3}{k\over 2}{\partial\over \partial \eta}|A_+^2|,
\end{equation}
where we only consider the amplified polarization and $\partial A_+/\partial \eta \simeq \sqrt{2\xi k a H }A_+$. By integrating from $k=0$ to $k_{\rm {max}}$, we arrive at
\begin{equation}
  \left \langle \EE\cdot\BB \right \rangle \propto e^{2\pi\xi} \left ({k_{\rm {max}}\over a}\right )^4 .
\end{equation}
For any finite range of comoving wave numbers, this contribution redshifts as $a^{-4}$.
However, in the presence of a slow rolling axion field $\phi$, new wave numbers are amplified all the time, meaning that $k_{\rm {max}}\simeq 2\xi aH$, leading to 
$  \left \langle \EE\cdot\BB \right \rangle \propto e^{2\pi\xi} H^4 $, which is constant in time.
Simply put, the largest wave numbers are dominating the backreaction. 

Let us now move to the case of the new dynamical attractor in the axion-$SU(2)$ system. The linearized equation for the gauge field fluctuations at late times can be approximated as
\begin{equation}
    T''_{R,L} + {2\over \eta^2 }m_Q \xi\,  T_{R,L} =0,
\end{equation}
where we ignored terms proportional to $T_{R,L}$ that grow slower than $\eta^{-2}$ at late times, as well as terms proportional to $\psi_{R,L}$ or $\psi'_{R,L}$, which are suppressed by slow-roll quantities. During the new attractor, $m_Q\xi\simeq -1$ (see equation \eqref{eqstage30}),
so the above equation has a growing late-time solution
$T_{R,L}\propto 1/\eta$, as seen for example in figure~\ref{pTRL_8panels}
(see also ref.~\cite{Adshead:2013nka} for a similar late-time evolution of $T_{R,L}$ in the original chromo-natural inflation model).

\begin{figure}
\centering
     \includegraphics[scale=0.6]{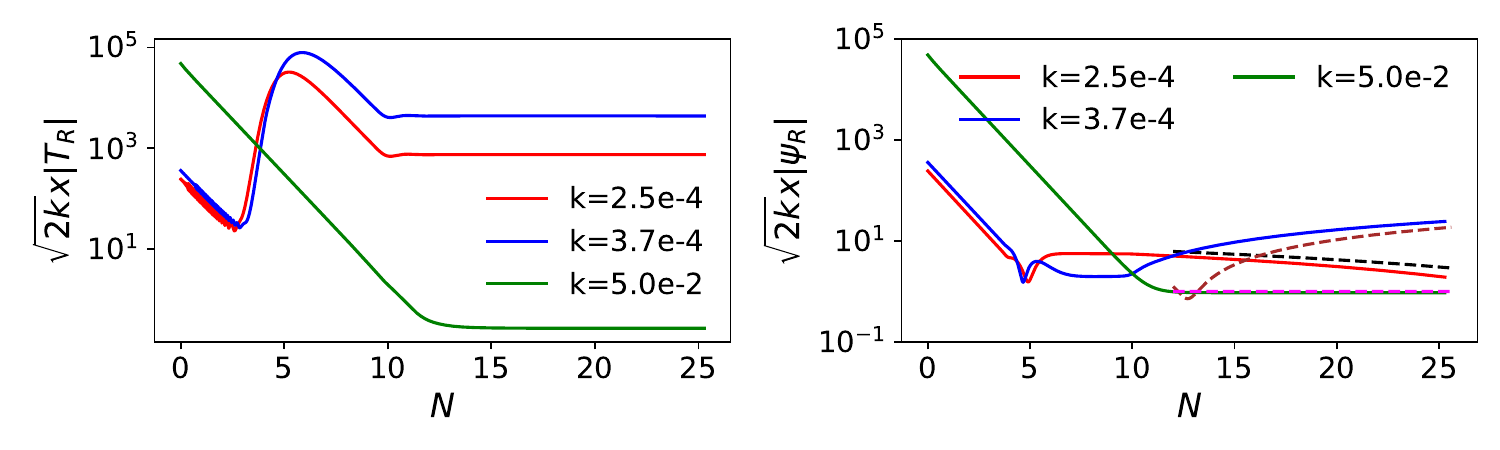}
\caption{The normalized mode-functions $\sqrt{2k}x|T_{R}|$ (left) and $\sqrt{2k}x|\psi_{R}|$ (right) as a function of $e$-folds $N$ for the run~$\mu4$ ($m_Q$=3.58) for different wave numbers.
The solid red, blue, and green curves correspond the numerical results for $k=2.5 \times 10^{-4}$, $3.7 \times 10^{-4}$, and $5.0 \times 10^{-2}$ respectively,
while the dashed black, brown, and magenta curves in the right panel represent the corresponding analytical results given in \Eq{analytical_psir} in the superhorizon limit.}
\label{comparison}
\end{figure}

The backreaction terms $ {\cal T}^Q_{\rm BR}, {\cal T}^{\chi}_{\rm BR}$ given in equations~\eqref{qbackint} and \eqref{chibackint} can be calculated similarly to
$\langle \EE\cdot \BB \rangle$ to scale as $ {\cal T}^Q_{\rm BR}, {\cal T}^\chi_{\rm BR} \propto k_{\rm {max}}$ for late times.
Thus, contrary to the AS solution, the new non-Abelian attractor does not require continuous amplification of new, ever-larger wave numbers.
Instead, it is supported by a fixed range of comoving wave numbers.
This is shown in figure~\ref{pcomp_sample_mu1_mu4_mu5}. 

Before we conclude, it is worth exploring the late-time superhorizon evolution of gravitational wave modes $\psi_{R}$ in the presence of the growing mode $T_{R}$. By keeping only the dominant terms of eq.~\eqref{psieqn} for $\eta\to 0$, we arrive at
\begin{equation}
   \psi_{R}''-\frac{2}{\eta^2}\psi_{R}=\frac{2 \sqrt{\epsilon_{Q_E}}}{\eta} T_{R}'+\frac{2 \sqrt{\epsilon_{Q_B}}}{\eta^2}m_QT_{R} 
   = \frac{2}{\eta^3} \left [
m_Q \sqrt{\epsilon_{Q_B}} - \sqrt{\epsilon_{Q_E}}
   \right
   ] T_{R}^{0} \, ,
\end{equation}
where we used $T_{R} = T_{R}^{0}/\eta$
at late times. 
This can be analytically solved to give \begin{equation}\label{analytical_psir}
    \eta \psi_{R} \sim {\rm const} - \frac{1+3\log(\eta)}{9}\left [
m_Q \sqrt{\epsilon_{Q_B}} - \sqrt{\epsilon_{Q_E}}
   \right
   ] T_{R}^{0} \, .\end{equation}
   We compare this analytic expression to the numerical results in figure~\ref{comparison}, showing excellent agreement.
We must note that $T_R^{0}$ depends on the wave number $k$, since different wave numbers experience different amplification, as shown in figure~\ref{comparison}.

\bibliographystyle{JHEP}
\bibliography{refs.bib}

\end{document}